\RequirePackage{lineno}
\documentclass[aps,prd,reprint,superscriptaddress,showpacs,floatfix,nofootinbib]{revtex4-1}

\usepackage{graphicx} 
\usepackage{dcolumn}  
\usepackage{bm}       
\usepackage{rotating}
\usepackage{amssymb}
\usepackage{amsmath}
\usepackage{multirow}

\def\cleoiii{CLEO~III}
\def\gev   {\mbox {GeV}}
\def\mev   {\mbox {MeV}}

\def\gevc  {\mbox {GeV/$c$}}
\def\mevc  {\mbox {MeV/$c$}}
\def\gevcc  {\mbox {GeV/$c^2$}}
\def\mevcc  {\mbox {MeV/$c^2$}}

\newcommand{\secref}[1]{Sec.~\ref{#1}}
\newcommand{\Secref}[1]{Section~\ref{#1}}

\newcommand{\figref}[1]{Fig.~\ref{#1}}
\newcommand{\Figref}[1]{Figure~\ref{#1}}
\newcommand{\tabref}[1]{Table~\ref{#1}}

\newcommand{\equationref}[1]{Eq.~(\ref{#1})}

\def \kskpi{\mbox{$K_S^0 K^\mp \pi^\pm \, $}}
\def \braf{\mbox{$\mathcal{B}_{K_S^0K\pi} \, $}}
\newcommand{\dtofav}{$D^0 \to K_S^0 K^- \pi^+$}
\newcommand{\dtosup}{$D^0 \to K_S^0 K^+ \pi^-$}
\newcommand{\dtokthreepi}{$D^0 \to K^- \pi^+ \pi^- \pi^+$}
\newcommand{\dtokpipio}{$D^0 \to K^- \pi^+ \pi^0$}
\def \cho{\mbox{$R_{K_S^0K\pi}$}} 
\def\spd{\mbox{$\delta^{K_S^0K\pi}_D$}}
\def \chokst{\mbox{$R_{K^*K}$}} 
\def\spdkst{\mbox{$\delta^{K^*K}_D$}}
\newcommand{\dzero}{$D^0$}
\newcommand{\dbar}{$\overline{D^0}$}
\def\Journal#1#2#3#4{{#1} {\bf #2}, #3 (#4)}

\def\NIMA{ Nucl. Instrum. Methods A}

\def\PLB{Phys. Lett.  B}
\def\PRL{Phys. Rev. Lett.}
\def\PRD{ Phys. Rev. D}

\begin{document}

\preprint{CPDRAFT 2011-10}
\preprint{CLNS 11/2081}   
\preprint{CLEO 11-07}

\title { \boldmath {Studies of the decays $D^0 \to K_S^0 K^- \pi^+$ and $D^0 \to K_S^0 K^+ \pi^-$  }}

\author{J.~Insler}
\author{H.~Muramatsu}
\author{C.~S.~Park}
\author{L.~J.~Pearson}
\author{E.~H.~Thorndike}
\affiliation{University of Rochester, Rochester, New York 14627, USA}
\author{S.~Ricciardi}
\affiliation{STFC Rutherford Appleton Laboratory, Chilton, Didcot, Oxfordshire, OX11 0QX, United Kingdom}
\author{C.~Thomas}
\affiliation{STFC Rutherford Appleton Laboratory, Chilton, Didcot, Oxfordshire, OX11 0QX, United Kingdom}
\affiliation{University of Oxford, Oxford OX1 3RH, United Kingdom}
\author{M.~Artuso}
\author{S.~Blusk}
\author{R.~Mountain}
\author{T.~Skwarnicki}
\author{S.~Stone}
\author{J.~C.~Wang}
\author{L.~M.~Zhang}
\affiliation{Syracuse University, Syracuse, New York 13244, USA}
\author{G.~Bonvicini}
\author{D.~Cinabro}
\author{M.~J.~Smith}
\author{P.~Zhou}
\affiliation{Wayne State University, Detroit, Michigan 48202, USA}
\author{T.~Gershon}
\affiliation{University of Warwick, Coventry CV4 7AL, United Kingdom}
\author{P.~Naik}
\author{J.~Rademacker}
\affiliation{University of Bristol, Bristol BS8 1TL, United Kingdom}
\author{K.~W.~Edwards}
\author{K.~Randrianarivony}
\affiliation{Carleton University, Ottawa, Ontario, Canada K1S 5B6}
\author{R.~A.~Briere}
\author{H.~Vogel}
\affiliation{Carnegie Mellon University, Pittsburgh, Pennsylvania 15213, USA}
\author{P.~U.~E.~Onyisi}
\author{J.~L.~Rosner}
\affiliation{University of Chicago, Chicago, Illinois 60637, USA}
\author{J.~P.~Alexander}
\author{D.~G.~Cassel}
\author{S.~Das}
\author{R.~Ehrlich}
\author{L.~Gibbons}
\author{S.~W.~Gray}
\author{D.~L.~Hartill}
\author{D.~L.~Kreinick}
\author{V.~E.~Kuznetsov}
\author{J.~R.~Patterson}
\author{D.~Peterson}
\author{D.~Riley}
\author{A.~Ryd}
\author{A.~J.~Sadoff}
\author{X.~Shi}
\author{W.~M.~Sun}
\affiliation{Cornell University, Ithaca, New York 14853, USA}
\author{J.~Yelton}
\affiliation{University of Florida, Gainesville, Florida 32611, USA}
\author{P.~Rubin}
\affiliation{George Mason University, Fairfax, Virginia 22030, USA}
\author{N.~Lowrey}
\author{S.~Mehrabyan}
\author{M.~Selen}
\author{J.~Wiss}
\affiliation{University of Illinois, Urbana-Champaign, Illinois 61801, USA}
\author{J.~Libby}
\affiliation{Indian Institute of Technology Madras, Chennai, Tamil Nadu 600036, India}
\author{M.~Kornicer}
\author{R.~E.~Mitchell}
\affiliation{Indiana University, Bloomington, Indiana 47405, USA }
\author{D.~Besson}
\affiliation{University of Kansas, Lawrence, Kansas 66045, USA}
\author{T.~K.~Pedlar}
\affiliation{Luther College, Decorah, Iowa 52101, USA}
\author{D.~Cronin-Hennessy}
\author{J.~Hietala}
\affiliation{University of Minnesota, Minneapolis, Minnesota 55455, USA}
\author{S.~Dobbs}
\author{Z.~Metreveli}
\author{K.~K.~Seth}
\author{A.~Tomaradze}
\author{T.~Xiao}
\affiliation{Northwestern University, Evanston, Illinois 60208, USA}
\author{D.~Johnson}
\author{S.~Malde}
\author{L.~Martin}
\author{A.~Powell}
\author{G.~Wilkinson}
\affiliation{University of Oxford, Oxford OX1 3RH, United Kingdom}
\author{D.~M.~Asner}
\author{G.~Tatishvili}
\affiliation{Pacific Northwest National Laboratory, Richland, WA 99352}
\author{J.~Y.~Ge}
\author{D.~H.~Miller}
\author{I.~P.~J.~Shipsey}
\author{B.~Xin}
\affiliation{Purdue University, West Lafayette, Indiana 47907, USA}
\author{G.~S.~Adams}
\author{J.~Napolitano}
\affiliation{Rensselaer Polytechnic Institute, Troy, New York 12180, USA}
\author{K.~M.~Ecklund}
\affiliation{Rice University, Houston, Texas 77005, USA}
\collaboration{CLEO Collaboration}
\noaffiliation


\date{\today}

\begin{abstract} 
The first measurements of the coherence factor \cho\ and the average strong--phase difference \spd\ in $D^0 \to K_S^0 K^\mp\pi^\pm$ decays are reported. These parameters can be used to improve the determination of the unitary triangle angle $\gamma$ in $B^-\to \widetilde{D}K^-$ decays, where $\widetilde{D}$ is either a \dzero\ or a \dbar\ meson decaying to the same final state, and also in studies of charm mixing. The measurements of the coherence factor and strong--phase difference are made using quantum--correlated, fully--reconstructed \dzero\dbar\ pairs produced in $e^+e^-$ collisions at the $\psi(3770)$ resonance. The measured values are \cho\ = 0.70 $\pm$ 0.08 and \spd\ = (0.1 $\pm$ 15.7)$^\circ$ for an unrestricted kinematic region and  \chokst\ = 0.94 $\pm$ 0.12 and \spdkst\ = (-16.6 $\pm$ 18.4)$^\circ$ for a region where the combined $K_S^0 \pi^\pm$ invariant mass is within 100 \mevcc\ of the $K^{*}(892)^\pm$ mass. These results indicate a significant level of coherence in the decay. In addition, isobar models are presented for the two decays, which show the dominance of the $K^*(892)^\pm$ resonance. The branching ratio $\mathcal{B}($\dtosup$)$/$\mathcal{B}($\dtofav$)$ is determined to be 0.592 $\pm$ 0.044 (stat.) $\pm$ 0.018 (syst.), which is more precise than previous measurements.

\end{abstract}
\pacs{13.25.Ft,12.15.Hh,14.40.Lb,11.80.Et}

\maketitle

\section{Introduction}

Decays of the $D^0$ meson to the \kskpi final state occur via a singly--Cabibbo--suppressed (SCS) transition. There are two possible decays, $D^0 \to K_S^0 K^- \pi^+$ and $D^0 \to K_S^0 K^+ \pi^-$ (charge conjugation is implied throughout the paper unless otherwise specified), where the first is decay is favored with respect to the second. These twin decays are of interest for a variety of reasons. Firstly when produced through a $B^- \to \widetilde{D} K^-$ decay, where $\widetilde{D}$ is either a $D^0$ or $\overline{D^0}$, they can be exploited to measure the Cabibbo--Kobayashi--Maskawa (CKM) phase $\gamma$~\cite{CKM}. Secondly a time--dependent study can be used to determine charm mixing parameters and probe for charm $CP$ violation~\cite{dmix}. Finally, examination of the resonant substructure of the decays allows for tests of SU(3) flavor symmetry in charm meson decays~\cite{jon}.

This paper describes several measurements performed with data collected by the CLEO experiment that are valuable for each of these purposes. These are a first determination of the coherence factor and average strong--phase difference for the two decays, a determination of the relative branching ratio of the decays that is more precise than previous measurements~\cite{PDG}, and the presentation of isobar models to describe each decay.
The coherence factor \cho\ and the average strong--phase difference \spd~\cite{coh} for $D^0 \to \kskpi$ are measured using quantum--correlated, fully--reconstructed
\dzero\dbar\ pairs produced in $e^+e^-$ collisions at the $\psi(3770)$
resonance. Knowledge of these parameters improves the
sensitivity of measurements of the unitary triangle angle $\gamma$ using $B^-$ meson decays to $\widetilde{D}K^-$ where the $\widetilde{D}$ decays to $K_S^0K^-\pi^+$ or $K_S^0K^+\pi^-$. To probe for the effects of physics beyond the standard model, it will be necessary to over--constrain the CKM quark--mixing matrix. An accurate measurement of $\gamma$ in $B\to \widetilde{D}K$ decays is of particular importance as the extracted phase is robust against new physics contributions, thus providing a standard model reference against which other measurements can be compared. The sensitivity to $\gamma$ comes from the interference of $b\to u $ and $b \to c $ quark transitions. The weak phase between these two transitions is $-\gamma$. The amplitudes are related by $\mathcal{A}(B^- \to \bar{D^0}K^-)/\mathcal{A}(B^- \to D^0K^-) = r_Be^{i(\delta_b-\gamma)}$, where $r_B$ has a value around 0.1 and is the absolute amplitude ratio, and $\delta_B$ is the strong--phase difference. Sensitivity to $\gamma$ is obtained by comparing the four separate decay rates involving $B^-$ and $B^+$, and the two possible $\tilde{D}$ final states. This method is directly analogous to the  ADS method~\cite{ADS}, originally proposed to exploit Cabibbo--favored (CF) and doubly--Cabibbo--suppressed (DCS) decays such as $K\pi$, $K\pi\pi\pi$, and $K\pi\pi^0$, rather than the SCS modes under consideration here. In the case of decays to multi--body flavor specific final states it is necessary to take into account that the amplitude ratio of \dtofav\ and \dtosup\ and the strong--phase difference will vary over phase space. Rate equations for different final states are given by~\cite{coh}
\begin{widetext}
\begin{eqnarray}
\label{simp}
\Gamma(B^{\mp} \to  {D}(K_S^0 K^{\mp}\pi^\pm)K^\mp) &\propto 1+ (r_Br_D^{K_S^0K\pi})^2 + 2r_Br_D^{K_S^0K\pi}R_{K_S^0K\pi}\cos(\delta_B - \delta_D^{K_S^0K\pi} \mp \gamma),\label{eq:gamfav}\\
\Gamma(B^{\mp} \to  {D}(K_S^0K^{\pm}\pi^\mp)K^\mp) & \propto (r_B)^2 + (r_D^{K_S^0K\pi})^2 + 2r_Br_D^{K_S^0K\pi}R_{K_S^0K\pi}\cos(\delta_B + \delta_D^{K_S^0K\pi} \mp \gamma),\label{eq:gamsup}
\end{eqnarray}

where $R_{K_S^0K\pi}$, $\delta_{K_S^0K\pi}$, and $r_D^{K_S^0K\pi}$ are defined as:
\begin {eqnarray}
R_{K_S^0K\pi}  e^{-i\delta_{K_S^0K\pi}} &=& \frac{\int \mathcal{A}^*_{K_S^0K^-\pi^+} (m^2_{K_S^0K},m^2_{K\pi}) \mathcal{A}_{K_S^0K^+\pi^-} (m^2_{K_S^0K},m^2_{K\pi}) dm^2_{K_S^0K}dm^2_{K\pi}}{A_{K_S^0K^-\pi^+}A_{K_S^0K^+\pi^-}}, \label{eq:defcho} \\
\mathrm{and\ } r_D^{K_S^0K\pi} &=& \frac{A_{K_S^0K^+\pi^-}}{A_{K_S^0K^-\pi^+}}\label{eq:defrd}.
\end{eqnarray}
\end{widetext}
Here, $ \mathcal{A}_{K_S^0K^\mp\pi^\pm} (m^2_{K_S^0K},m^2_{K\pi})$ is the amplitude for $D^0 \to K_S^0 K^\mp\pi^\pm$ at a point in multi--body phase space described by the square of the invariant mass of the $K_S^0K$ and $K\pi$ pairs, ($m^2_{K_S^0K}$, $m^2_{K\pi})$ respectively, and $A^2_{K_S^0K^\mp\pi^\pm} = \int | \mathcal{A}_{K_S^0K^\mp\pi^\pm} (m^2_{K_S^0K},m^2_{K\pi})|^2 dm^2_{K_S^0K}dm^2_{K\pi}$.  
Analogous equations can be written for the decays \dtokthreepi\ and \dtokpipio\ and the coherence factors and average strong--phase differences for these final states have already been measured~\cite{jim}. 
One difference between the $K^\mp\pi^\pm$, $K^\mp\pi^\pm\pi^+\pi^-$ and $K^\mp\pi^\pm\pi^0$ modes and the \kskpi\ mode is that the value $r_D^{K_S^0K\pi}$ is approximately ten times larger than the corresponding parameters for the other decays. The larger value for $r_D^{K_S^0K\pi}$ occurs because both final states occur via SCS transitions, while in the case of the $K^\mp\pi^\pm$, $K^\mp\pi^\pm\pi^+\pi^-$ and $K^\mp\pi^\pm\pi^0$ decay modes one transition is CF and the other DCS. The larger value of $r_D$ means that the per--event sensitivity to $\gamma$ from the \dtofav\ decays [\equationref{eq:gamfav}] is increased in comparison to $D^0\to K^-\pi^+$ decays, while the per--event sensitivity to $\gamma$ from the \dtosup\ decays [\equationref{eq:gamsup}] is reduced in comparison to $D^0 \to K^+\pi^-$ decays. The overall sensitivity to $\gamma$ remains significant if the coherence is large. Therefore knowledge of the coherence factor in $D^0 \to \kskpi$ decays is valuable, as it enables these channels to be added to the set of modes with which $\gamma$ can be measured using $B^- \to \widetilde{D} K^-$ decays. The coherence factor and average strong--phase difference are also valuable for time--dependent charm--mixing measurements as described in~\cite{dmix}. 

In a region around a resonance that is prominent in both decays, the coherence is expected to be close to one. A measurement in a bin of the Dalitz plot around such a resonance is therefore also useful as the higher coherence improves the per--event sensitivity in measurements to determine $\gamma$. In the case of $D^0\to \kskpi$ decays a prominent intermediate resonance is the $K^{*}(892)^\pm$. Therefore, in this paper the measurement of the coherence factor and the average strong--phase difference is performed for both an unrestricted kinematic region and also for events where the $K_S^0\pi^\pm$ invariant mass is within 100 \mevcc\ of the nominal $K^{*}(892)^\pm$ mass~\cite{PDG}. The value of 100 \mevcc\ corresponds approximately to twice the natural width of the $K^*(892)^{\pm}$ meson.

In addition to the measurement of the coherence factor and average strong--phase difference, two further studies are also presented. First, there is the determination of the isobar models for the two decays. These can be used to test the predictions of SU(3)--flavor symmetry in charm meson decays, in particular the relative amplitudes and phases of the modes $D^0 \to K^{*0}\overline{K^0}$ and $D^0 \to \overline{K^{*0}}K^0$~\cite{jon}. The isobar models are used to estimate systematic uncertainties for the other measurements presented here and will also be of use as an input to simulation when performing future measurements that exploit $D^0 \to \kskpi$ decays.  Secondly, the ratio of the branching fractions for the suppressed with respect to the favored decay, \braf$=\mathcal{B}($\dtosup$)$/$\mathcal{B}($\dtofav$)$, is measured. The current world average is 0.79 $\pm$ 0.19~\cite{PDG}, and a more accurate determination is possible using CLEO data. This is valuable as a better measurement of \braf allows for a corresponding improvement in the knowledge of $r_D^{K_S^0K\pi}$, which is an important input to the coherence factor analysis, and also to measurements of $\gamma$ and studies of charm mixing. The isobar model studies are based on data collected both at $\sqrt{s}=3770$ \mev\ and from charm mesons produced in the continuum at higher energies; the branching ratio measurement exploits only the latter sample.

This paper is organized as follows. ~\Secref{sec:datasamp} gives an overview of the detector configurations used and the data samples that are used for all the studies presented in this paper. ~\Secref{sec:ampdefs} describes the isobar models and ~\secref{sec:ratio} details the measurement of the ratio of branching fractions. ~\Secref{sec:coh} describes the measurement of the coherence factor and the strong--phase difference; and in addition the sensitivity to $\gamma$ using these results is discussed. A summary is given in ~\secref{sec:sum}.

\section{Data samples}
\label{sec:datasamp}

Two different data samples collected by the \cleoiii\ and CLEO-c detectors are used for the studies in this paper. A total of 15.3 fb$^{-1}$ of data was collected by the \cleoiii\ detector. For these data the Cornell electron storage ring (CESR) accelerator operated at a range of energies from $\sqrt{s}=7.0$ \gev\ to $\sqrt{s}=11.2$ \gev. For the data collected by the CLEO-c detector analyzed in this paper, the CESR energy was $\sqrt{s}=3770$ \mev\ and an integrated luminosity of 0.818 fb$^{-1}$ of data was accumulated.  The CLEO~III/CLEO-c detector was a solenoidal detector which included a gaseous tracking system for the measurement of charged particle momenta and ionization energy loss, a ring--imaging Cherenkov (RICH) detector to aid in particle identification, and a CsI crystal calorimeter to measure the energy of electromagnetic showers. The tracking system was particularly important in defining the candidate signal sample for the measurements in this paper. The \cleoiii\ tracking system ~\cite{track} consisted of a silicon strip vertex detector and a large drift chamber. The silicon was replaced by a six--layer wire vertex detector in the CLEO-c configuration ~\cite{track2}. The trackers achieved charged particle momentum resolution of 0.35$\%$ (0.6$\%$) at 1 \gevc\ in the 1.5 T (1.0 T) axial magnetic field used in \cleoiii\ (CLEO-c) detector, respectively. The detectors are described in detail elsewhere~\cite{III}.

The intermediate resonances contributing to the \dtofav\ and \dtosup\ decays are not identical, so in constructing isobar models it is necessary to know the flavor of the decaying \dzero. The isobar models are determined from flavor--tagged data from the \cleoiii\ and CLEO-c detectors. The data collected by \cleoiii\ are flavor--tagged using the decay $D^{*+} \to D^0 \pi^{+}$, in which the flavor of the decaying neutral $D$ meson is inferred from the charge of the pion. In the case of the data collected by the CLEO-c detector the interaction of interest is $e^+e^- \to \psi(3770) \to D^0\bar{D^0}$ where one $D$ meson, the signal side, decays to \kskpi\ and the other, the tag side, decays to either $K^\mp\pi^\pm$, $K^\mp\pi^\pm\pi^+\pi^-$, or $K^\mp\pi^\pm\pi^0$. Reconstruction of both $D$ mesons is referred to as a double tag. The flavor of the signal side $D$ meson is inferred from the charge of the kaon on the tag side, taking advantage of the fact that the CF transition, for example $D^0 \to K^- \pi^+$, is dominant. Corrections to this assumption from DCS decays are considered in the assignment of systematic uncertainties. The measurement of \braf\ is performed using only the flavor--tagged data from \cleoiii.

 The coherence factor measurement uses data only from CLEO-c. To maximize the sensitivity to the coherence factor a range of double tags are used. These include the flavor tags used for the isobar model and also double tags where the tag side $D$ meson decays to a $CP$ eigenstate or $K^0_{S,L}\pi^+\pi^-$, which are states of mixed $CP$.  Table~\ref{tab:states} lists all the reconstructed double tags used for the coherence factor measurement. The neutral mesons in the final states are reconstructed via the following decays: $\pi^0 \to \gamma\gamma$, $K_S^0\to \pi^+\pi^-$, $\omega \to \pi^+\pi^-\pi^0$, $\eta\to\gamma\gamma$, $\eta \to \pi^+\pi^-\pi^0$, and $\eta' \to \eta(\gamma\gamma)\pi^+\pi^-$. The $K_L^0$ meson is not reconstructed directly. Its presence is inferred via a missing--mass technique as described in~\secref{sec:datacleoc}. The rest of this section describes the selection requirements placed on all the different data types used to perform the measurements presented in this paper.    
\begin{table}[htbp]
\begin{center}
\caption[]{A list of the CLEO-c double tags that are selected, split by the category of the tag side $D$ meson. All the listed double tags are used in the coherence factor measurement, but only the flavor double tags are used in the isobar model analysis. \label{tab:states}}
\begin{tabular}{ll}\hline \hline
Tag group & Tag side decays\\ \hline
Flavor & $K^\pm\pi^\mp$, $K^\pm\pi^\mp \pi^0$,  $K^\pm\pi^\mp\pi^+ \pi^-$\\
\multirow{2}{*}{$CP$ even} & $K^+K^-$, $\pi^+\pi^-$, $K_L^0 \pi^0$, $K_L^0 \eta$, \\
& $K_L^0 \omega$, $K_L^0 \eta'$, $K_S^0 \pi^0\pi^0$ \\
$CP$ odd &  $K_S^0 \pi^0$, $K_S^0 \eta$, $K_S^0 \omega$, $K_S^0 \eta'$, $K_L^0 \pi^0 \pi^0$ \\
Mixed $CP$ & $K_S^0 \pi^+ \pi^-$, $K_L^0 \pi^+ \pi^-$\\ \hline \hline
\end{tabular}
\end{center}\end{table}

\subsection{Data from \cleoiii}
\label{sec:datacleo3}

For the flavor--tagged $D^*$ data from \cleoiii, various requirements are made on properties of charged tracks, composite particles and mass or vertex fits in order to achieve samples of $ D^0 \to K_S^0 K^\pm \pi^\mp$ with high purity. The kaon and pion from the $D$ meson and the tagging pion from the $D^*$ are required to pass close to the interaction point: within 5 cm along the beam axis, and 5 mm perpendicular to it. The cosine of the angle between the track and the beam pipe is required to lie between $-0.9$ and 0.9, to ensure that the track lies within the detector acceptance. The momentum of the slow pion track is required to be between 0.15 and 0.50 \gevc. All other tracks have a minimum momentum requirement of 0.2 \gevc\ and a maximum momentum of 5.0 \gevc. Charged pions are used to reconstruct the $K_S^0$ candidates and the flight significance of the $K_S^0$ candidate is required to be greater than 8, where the flight significance is given by the measured flight distance divided by the measured flight distance uncertainty. Kaons and pions are distinguished by using information from the RICH detector if the momentum of the kaon track is above 500 \mevc, otherwise ionization energy loss from the drift chamber is used. Kaon and pion candidates are combined with a $K_S^0$ candidate to form a $D^0$ candidate. The slow pion candidate is combined with the $D^0$ candidate to form a $D^*$ candidate. In order to suppress combinatoric backgrounds and backgrounds from $B$ meson decays, the momentum of the $D^*$ is required to be greater than half its maximum possible value (for a given beam energy). Loose selection requirements are placed on the $D^0$ and $D^*$ fitted vertex $\chi^2$ to remove combinatoric candidates.

\begin{figure*}[htbp]
\begin{center}
 \includegraphics[width=0.99\textwidth]{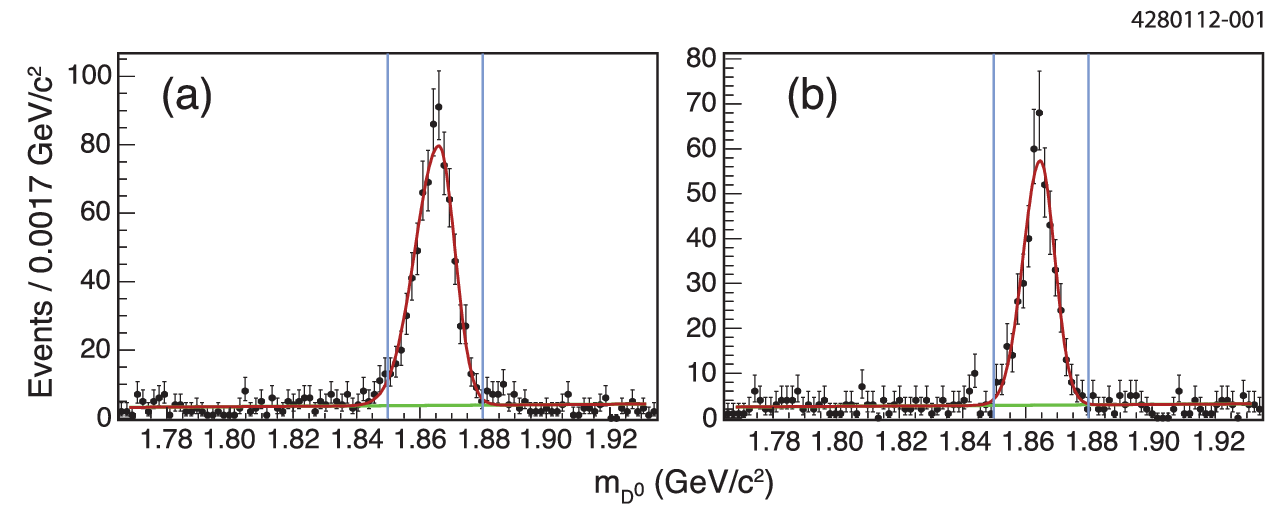}
\end{center}
\caption{The reconstructed mass distributions (points) of (a) \dtofav\ and (b) \dtosup\ in \cleoiii\ data. The signal window selection in $\Delta m_D$ has been applied. The total fit projection (dark line) and background (light line) component are also shown. The signal window in the reconstructed mass variable is indicated by the vertical lines.  \label{fig:c3data}}
\end{figure*}

Signal candidates are chosen using the reconstructed $D^0$ meson mass $m_{D^0}$ and $\Delta m_D = m_{D^*} - m_{D^0}$, where $m_{D^*}$ is the reconstructed $D^*$ meson mass.  The signal region encloses a region 15 MeV/$c^2$ either side of the nominal $D^0$ meson mass, and 1 \mevc$^2$ either side of the nominal mass difference, where the nominal $D^0$ and $D^*$ masses are taken from ~\cite{PDG}. These ranges are picked as they correspond approximately to a three standard deviation window in either the $D^0$ mass or $\Delta m_D$ projection, where the width of the signal is determined by a fit to each distribution. In each case, when the fit is performed on one variable the signal box cut has been applied to the other variable. The signal $m_{D^0}$ distribution is described by a bifurcated Gaussian and the background is described as a first--order polynomial. In $\Delta m_D$ the signal is also described by a bifurcated Gaussian. The background shape is parametrized by $(\Delta m_D/a)^2[1-e^{-(\Delta m_D-d)/c} + b(\Delta m_D/d -1)]$ where $a$, $b$, $c,$ and $d$ are parameters determined by the fit. A correction is required for \dtofav\  decays as Monte Carlo simulation studies indicate the presence of a peaking background. This background is the $D^0 \to K^-\pi^+\pi^+\pi^-$ decay misidentified as $K_S^0K^-\pi^+$ and contributes approximately 2$\%$ to the peak. Monte Carlo simulation studies indicate no other significant peaking backgrounds for either decay. The signal yields from the fits are corrected to remove the $D^0 \to K^-\pi^+\pi^+\pi^-$ contribution. The purities of the samples in the signal region are evaluated by averaging the purity determined from the fits in both distributions, which are consistent with each other. The purity of the $D^0 \to K_S^0 K^- \pi^+$ sample is (92.4 $\pm$ 0.5)$\%$ with a signal yield of 640 $\pm$ 27 in the signal region after background subtraction. The purity of the  $D^0 \to K_S^0 K^+ \pi^-$ sample is (90.0 $\pm$ 0.7)$\%$ with a signal yield of 406 $\pm$ 24 in the signal region. The given uncertainties on the purities and signal yields are statistical only. \Figref{fig:c3data} shows the $D^0$ mass distribution and fit projections for favored and supressed decays after the signal window in $\Delta m_D$ has been applied. The signal yield determination is also performed in the restricted kinematic region where the signal $K_S^0$ and $\pi^\pm$ have an invariant mass within 100 \mevcc\ of the nominal $K^{*}(892)^\pm$ resonance. In this restricted region the signal yield is  445 $\pm$ 22 for \dtofav\ events and 166 $\pm$ 14 for \dtosup\ events, and the purities are (93.8 $\pm$ 0.9)$\%$ and (90.9 $\pm$ 1.2)$\%$ respectively.

\subsection{Data from CLEO-c}
\label{sec:datacleoc}

A variety of double tags, as listed in ~\tabref{tab:states}, are reconstructed in the CLEO-c data. The standard CLEO-c selection criteria for $\pi^+$, $\pi^0$ and $K^+$ mesons are adopted and are described in~\cite{selec}. In addition to these criteria, the $K_S^0$ vertex is required to be separated from the interaction point by at least two standard deviations and candidate $K_S^0 \to \pi^+\pi^-$ decays are required to have a mass within 7.5 MeV/$c^2$ of the nominal mass. This criterion is reduced to within 5.0 MeV/$c^2$ of the nominal mass when the $K_S^0$ forms part of the $\kskpi$ signal tag. This tighter cut is to reduce background from the $D^0\to K^-\pi^+\pi^+\pi^-$ decay faking the \dtofav\ signal. The $\eta \to \gamma\gamma$ candidates are reconstructed in a similar fashion to $\pi^0$ candidates with the requirement that the invariant mass is within 42 \mevc$^2$ of the nominal mass; the same requirement is applied to $\eta \to \pi\pi\pi^0$ candidates. Candidates for $\omega \to \pi\pi\pi^0$ decays are required to be within 20 MeV$/c^2$ of the nominal $\omega$ mass. The $\eta' \to \eta \pi\pi$ candidates are required to have an invariant mass in the range 950 to 964 MeV/$c^2$. All nominal masses are taken from ~\cite{PDG}.

\begin{figure*}[htbp]
\begin{center}
 \includegraphics[width=0.99\textwidth]{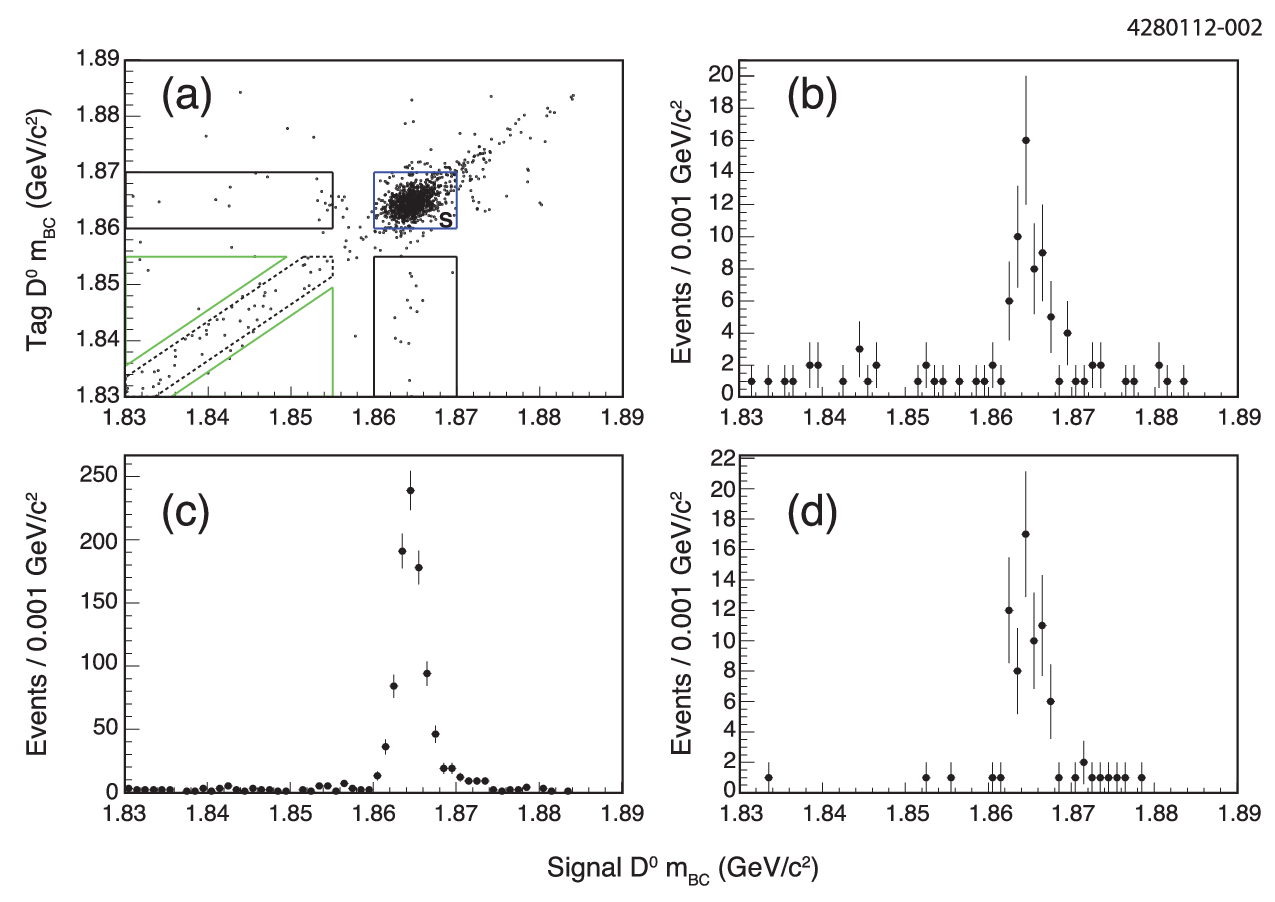}
\end{center}
\caption{CLEO-c data distribution of fully--reconstructed double tags on the $[m_{bc}(D_{\mathrm{sig}}), m_{bc}(D_{\mathrm{tag}})]$ plane (a) where the signal region is labeled S and the other indicated boxes are the sideband regions. The projections of the $m_{bc}(D_{\mathrm{sig}})$ are shown separately for candidates that are tagged by (b) $CP$ eigenstates, (c) the flavor states, and (d) $K_S^0\pi^+\pi^-$. \label{fig:kpimbc}
}
\end{figure*}

For final states that do not contain a $K_L^0$, two kinematic variables are considered: the beam--constrained candidate mass $m_{bc} = \sqrt{s/(4c^4) - \mathbf{p}^2_D/c^2}$, where $\mathbf{p}_D$ is the $D$ candidate momentum, and $\Delta E = E_D - \sqrt{s}/2$, where $E_D$ is the sum of the energies of the daughters that comprise the $D$ meson candidate. Decays with correctly identified signal and tag decays peak at the nominal $D^0$ mass in $m_{bc}$ and zero in $\Delta E$. Mode--dependent requirements are placed on the signal and tag candidates such that $\Delta E$ is less than three standard deviations from zero. The double--tag yield is determined by counting events in signal and sideband regions of the $[m_{bc}(D_{\mathrm{sig}}), m_{bc}(D_{\mathrm{tag}})]$ plane. ~\Figref{fig:kpimbc} shows all the fully--reconstructed double tags on the $[m_{bc}(D_{\mathrm{sig}}), m_{bc}(D_{\mathrm{tag}})]$ plane with the signal region marked S, and the sidebands illustrated. The rectangular sidebands are populated by decays where only one of the $D$ mesons has been correctly identified. The triangular sidebands are sparsely populated by combinatoric background as neither $D$ meson candidate has a mass consistent with a $D$ meson. The remaining narrow diagonal sideband is populated with two types of events. One of these is continuum background where the original interaction was $e^+e^- \to q\bar{q}$, where $q=$$u$, $d$, or $s$. The other type of background present in this sideband is where tracks have been assigned to the wrong $D$ meson. Also shown in ~\figref{fig:kpimbc} are the projections of the signal $D$ meson mass in different tag groups.

The different sidebands contain contributions from distinct types of combinatorial background. The yields in these sidebands are scaled and subtracted from the yield in the signal region. There are some backgrounds that peak in the signal region and hence cannot be determined from the sidebands. These peaking background contributions are estimated from Monte Carlo samples. {\sc EVTGEN}~\cite{evtgen} is used to generate the decays and {\sc GEANT}~\cite{GEANT} is used to simulate the CLEO-c detector response. A generic sample of Monte--Carlo--simulated data of $\psi(3770) \to D \overline{D}$ decays, with a size corresponding to an integrated luminosity twenty times that of data is used. The subsequent decay of the $D^0$ or $D^+$ meson is to all known decay channels. A further sample of simulated continuum data of $e^+e^-\to q\overline{q}$ where $q$ is either the $u$, $d$, or $s$ quark, with a size corresponding to an integrated luminosity five times that of the data is also analyzed.  All signal events, determined from truth information, are removed, and all background events that pass the selection criteria for any double tag are used to determine the peaking background for that double tag. The background contribution from the continuum is determined to be small in comparison to that from generic $D\overline{D}$ decays. The samples are combined by scaling the small continuum contribution to the integrated luminosity size of the generic sample. The statistical uncertainty due to the Monte Carlo background yields is very small in comparison to the uncertainty on the data yields. The uncertainty on the scale factor used to determine the peaking background yields from Monte Carlo simulation is 10$\%$. This is determined by comparing all the double tag sideband yields in data to those in Monte Carlo to account for differences between Monte Carlo simulation and data.

 The largest peaking background to \dtofav\ decays that remains after selection is from \dtokthreepi\ decays where a $\pi^+\pi^-$ pair forms a $K_S^0$ candidate.  The background--subtracted signal region yields and purities from flavor tags are given in ~\tabref{tab:pseudoyields} where the errors include the statistical uncertainties from the signal, sideband and Monte Carlo yields. Yields of other fully--reconstructed $CP$ double tags and the $K_S^0\pi\pi$ tag are given in~\tabref{tab:CPyields}. For the $K_S^0\pi^+\pi^-$ tag, the yields are required separately for 16 specific regions on the $K_S^0\pi^+\pi^-$ Dalitz plot. These regions are the equal phase--difference $\Delta\delta_D$ bins reported in~\cite{kspipipap}, which are defined using the isobar model reported in ~\cite{babar}. Signal yields in these bins are determined and range between 0.7 $\pm$~1.0 and 7.8$\pm$~2.8. These uncertainties include the statistical uncertainties on the signal region yield and the subtracted background.
\begin{table}[htbp]
\begin{center}
\caption[]{ Background--subtracted yields and purities of $D^0 \to \kskpi$ for each of the flavor tags in the CLEO-c data. The uncertainties include the statistical uncertainties of the yields in the signal region, sideband regions and Monte Carlo yields that determine the peaking backgrounds. \label{tab:pseudoyields}}
\begin{tabular}{lcccc}\hline \hline
Tag & \multicolumn{2}{c}{\dtofav\  }& \multicolumn{2}{c}{\dtosup\ }\\ 
& Yield & Purity $\%$& Yield & Purity$\%$\\ \hline
$K \pi$ & 122.0$\pm$11.5 & 93.2$\pm$0.5 & 80.6$\pm$9.0 & 99.5$\pm $0.1\\
$K \pi\pi\pi$ &161.8$\pm$13.4 & 89.8$\pm$0.5 & 101.5$\pm$10.2 & 97.6$\pm $0.3 \\
$K \pi\pi^0$ & 221.4$\pm$15.6 & 91.0$\pm$0.4 & 120.8$\pm$11.1 & 97.4$\pm $0.2\\
 \hline \hline
\end{tabular}
\end{center}\end{table}

\begin{table*}[htpb]
\begin{center}
\caption{Background--subtracted yields for fully--reconstructed $CP$ tags and $K_S^0\pi\pi$ tag over all Dalitz plot bins. The uncertainties include the statistical uncertainties of the yields in the signal region, sideband regions and Monte Carlo yields that determine the peaking backgrounds. \label{tab:CPyields}}
\begin{tabular}{lcccccccccc}
\hline \hline
 & $KK$  & $\pi\pi$  & $K_S^0\pi^0\pi^0$  & $K_S^0\pi^0$  & $K_S^0\omega$  & $K_S^0\eta ( \gamma\gamma )$  & $K_S^0\eta(\pi\pi\pi^0)$  & $K_S^0\eta '$ & $K_S^0\pi\pi$ \\
\hline
Yield & 1.7$\pm$2.5 & 3.2$\pm$3.1 & 10.6$\pm$3.5 & 22.8$\pm$4.9 & 12.3$\pm$3.8 & 6.9$\pm$2.6 & 0.4$\pm$1.0 & 0.8$\pm$1.0 & 60.7$\pm$8.1\\
\hline \hline
\end{tabular}
\end{center}
\end{table*}

\begin{figure}[htbp]

\begin{center}
{\includegraphics[width=0.99\columnwidth]{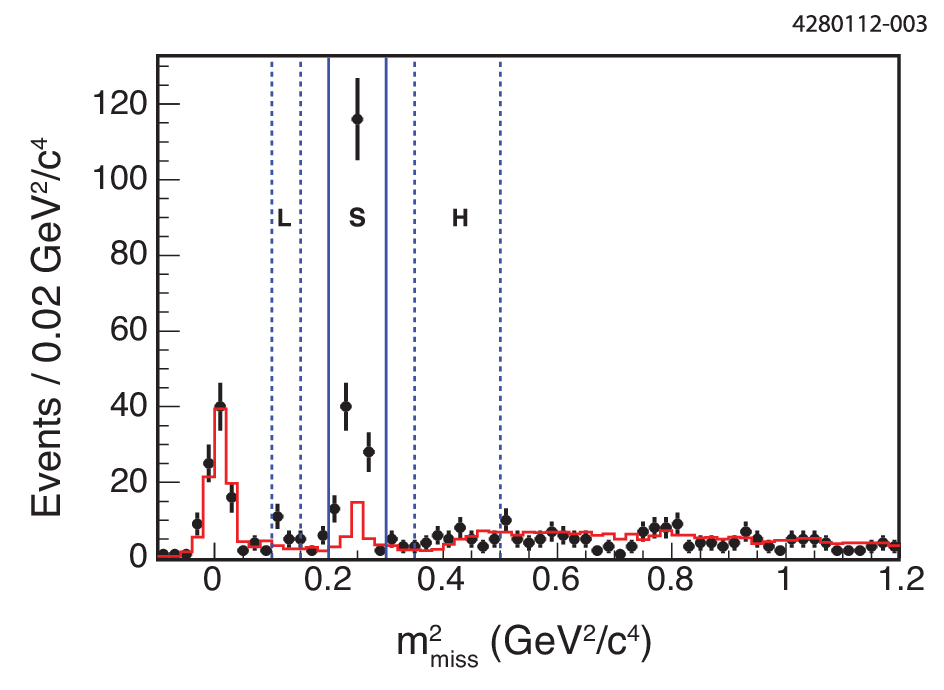}}
\caption{The distribution for the missing mass squared for the $K_L^0\pi^+\pi^-$ tagged candidates. The points show the data and the solid line is the Monte Carlo background prediction, scaled to the data luminosity. The region between the two straight solid lines is the signal region, marked S, and the regions between the two dotted lines are the sidebands, marked L and H, used to determine the combinatorial background. \label{fig:klstates}}
\end{center}
\end{figure}

For final states that include a $K_L^0$ meson, the missing mass squared recoiling against the fully--reconstructed $D$ candidates and the particles from the other $D$ decay containing the $K_L^0$ meson is computed. Events consistent with the mass of the $K_L^0$ meson squared are selected. This technique was introduced in~\cite{Klreco}. Further selection requirements are as those listed in~\cite{kspipipap}. The combinatoric background yield in the signal region is estimated from the population in the lower (L) and upper (H) missing mass squared sidebands. Information from the generic simulation is used to determine the relative composition of the sidebands and the signal region to estimate better the combinatorial background. There are also peaking backgrounds that only appear in the signal region of the missing mass squared distribution. The largest contribution of peaking background arises from $D^0 \to K_S^0 X$ decays, where the $K_S^0$ is not reconstructed and the decay is mistaken for  $D^0 \to K_L^0 X$. As an example,~\figref{fig:klstates} shows the missing mass squared spectrum for the $K_L^0\pi^+\pi^-$ tag in data and the background Monte Carlo distribution scaled to the integrated luminosity of the data. In the signal region the peaking contribution of the $K_S^0\pi^+\pi^-$ decays can be seen in the background Monte Carlo. The lower and upper sidebands are chosen to avoid any peaking structures that do not contribute to the background under the peak such as the contribution of $D^0 \to K^-\pi^+\pi^0$ that is observed to the left of the lower mass sideband in ~\figref{fig:klstates}.

 The background--subtracted signal yields of the $CP$ tags with a $K_L^0$ meson in the tag side and the $K_L^0\pi\pi$ tag are given in~\tabref{tab:KLyield}. For the $K_L^0\pi\pi$ tag, the candidate events are separated into regions on the Dalitz plot and the background--subtracted yields in each region range from $0.8\pm1.4$ to $24.8\pm 5.7$. The purity of $K_L^0\pi^0$ tagged candidates is 67$\%$. Given the very low yield for the candidates tagged by $K_L^0 \eta (\gamma \gamma)$ and the lower branching fraction and reconstruction efficiency for the $K_L^0\eta(\pi^+\pi^-\pi^0)$ tag, the $K_L^0\eta(\pi^+\pi^-\pi^0)$ tag is not worth pursuing. Although several of the yields are close to zero the corresponding tag categories are retained in the analysis as the sensitivity to the coherence factor comes from the combination of many tags.
The selection is repeated for all tags where the invariant mass of the $K_S^0$ and $\pi$ from the signal decay is required to lie within 100 \mevcc\ of the $K^*(892)^\pm$ nominal mass. The yields in this restricted region typically lie between 40$\%$ and 80$\%$ of the reported unrestricted yields, depending on tag type. The selected samples in the restricted region have similar purities to those in the unrestricted region. The efficiency--corrected yields of these samples are given in \secref{sec:cohres}.

\begin{table*}[thpb]
\begin{center}
\caption{Background--subtracted signal yields for the $CP$ tags with a $K_L^0$ and  the $K_L^0\pi\pi$ tag over all Dalitz plot bins. The uncertainties include the statistical uncertainties of the signal region and subtracted background estimates.\label{tab:KLyield}}
\begin{tabular}{lcccccc}
\hline \hline
 & $K_L^0\pi^0$  & $K_L^0\omega$  & $K_L^0\eta(\gamma\gamma)$  & $K_L^0\eta '$  & $K_L^0\pi^0\pi^0$ & $K_L^0\pi\pi$ \\
\hline
Yield & 22.0$\pm$5.8 & 2.9$\pm$2.7 & $-$0.1$\pm$1.8 & 0.3$\pm$1.0 & 2.8$\pm$2.0 & 150.9$\pm$14.2\\
\hline \hline
\end{tabular}
\end{center}
\end{table*}

\section{Isobar Models}

\label{sec:ampdefs}
\begin{figure*}[htbp]
\begin{center}
     \includegraphics[width=0.9\textwidth]{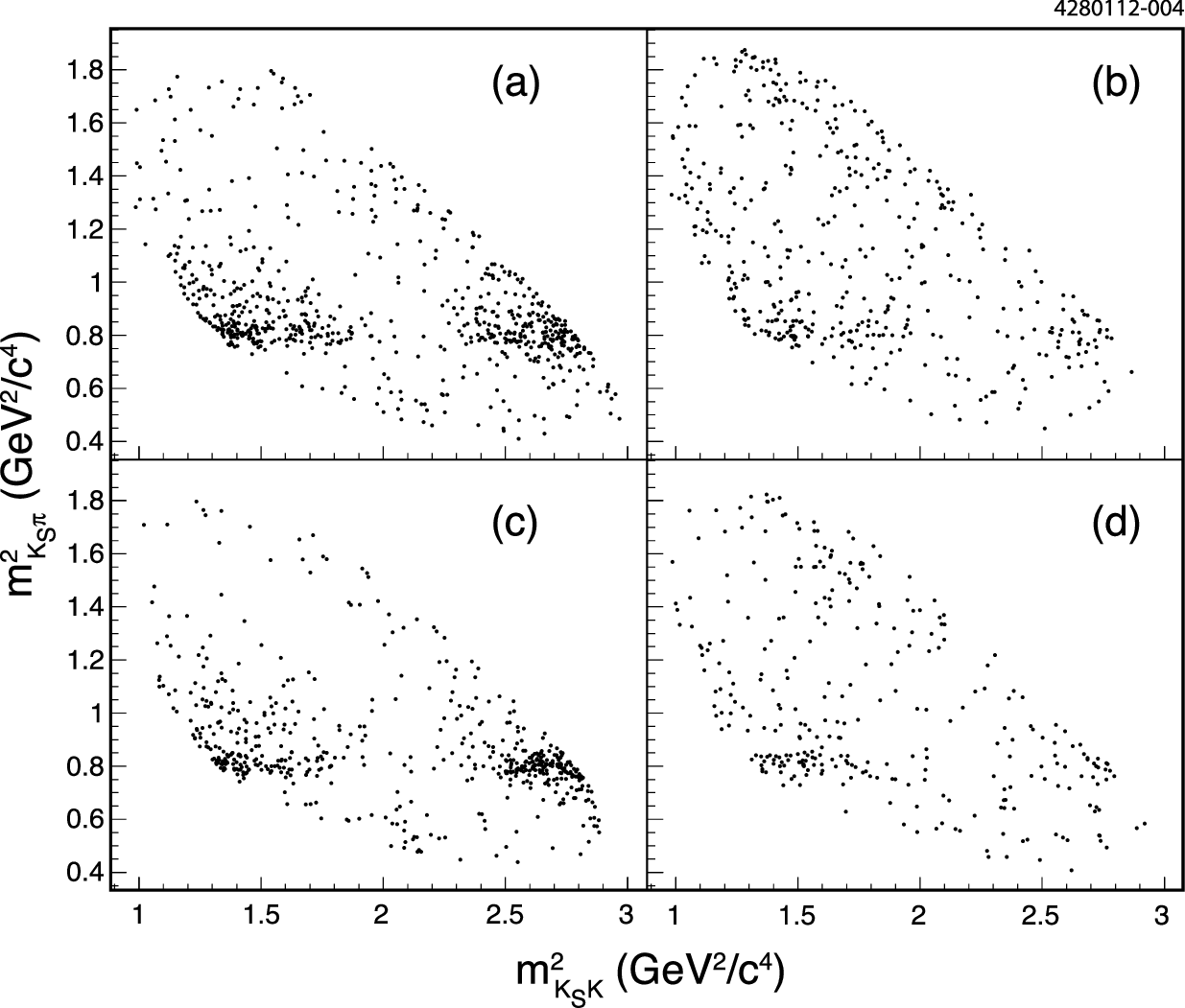}
\end{center}
\caption{Dalitz plot distributions for (a)[(c)] \dtofav\ and (b)[(d)] \dtosup\ for flavor--tagged CLEO~III [CLEO-c] data.\label{fig:dalcomb}}
\end{figure*}

The intermediate resonance structure in the decays \dtofav and \dtosup can be visualized through the distribution of the data over the Dalitz plot. The Dalitz--plot distributions for each flavor--tagged data set are shown in ~\figref{fig:dalcomb} where all three flavor tags used in CLEO-c are plotted together. The horizontal band in each plot is a clear indication of the presence of an intermediate $K^*(892)^\pm$ resonance. The remainder of this section describes the construction of isobar models for the favored and suppressed modes that parametrize the relative amplitudes and phases of all intermediate resonances that contribute to the decay.

\subsection{Signal model for the amplitude fit}
\label{sec:ampdefsdet}

The signal isobar model considers the decay $D^0 \to C R, R\to AB$, where $A, B, C$ are
pseudoscalar final state particles, and $R$ is an intermediate resonance
with spin~0,~1, or~2. The corresponding four--momenta are $p_D$ for the
initial $D$, $p_R$ for the resonance, and $p_A, p_B, p_C$ for the final
state particles. For the invariant mass of the resonance (which, due
to its width, will vary event by event), the symbol $m_{AB}$ is used, with
$m_{AB}^2 = p_R^2 = (p_A + p_B)^2$. For the ``nominal'' mass of the
resonance the symbol $m_R$ is used.
The matrix element for a $D^0 \to C R, R\to AB$ decay chain is
of the general form for resonance $R_i$
\begin{equation}
\label{eq:fullmi}
 \mathcal{M}_i = F_D
 \frac{\Omega_J}{m_R^2 - m_{AB}^2 - i m_R \Gamma_{AB}}
  F_R,
\end{equation}
where $F_D$ and $F_R$ are orbital angular momentum barrier penetration factors,
$1/(m_R^2 - m_{AB}^2 - i m_R \Gamma_{AB})$ is the line--shape of the
resonance (in this example Breit--Wigner; other choices are possible),
and $\Omega_J$ is the spin factor for a resonance of spin $J$. The running width, $\Gamma_{AB}$ is given by
\begin{equation}
\Gamma_{AB} = \Gamma_r \left(\frac{p_{AB}}{p_r}\right)^{2J+1} \left(\frac{m_R}{m_{AB}}\right)F_R^2,
\end{equation}
where $\Gamma_r$ is the natural width, $p_{AB}$ is the momentum of one of the daughters in the $AB$ rest frame, and $p_r$ is the momentum of one of the daughters in the resonance rest frame.
  The spin factors are calculated as described in~\cite{pene}, and are
\begin{widetext}
\begin{eqnarray}
\Omega_0 &=& 1,\\
\Omega_1 &=& (p_D^{\mu} + p_C^{\mu}) T_{\mu\alpha}(p_B^{\alpha} - p_A^{\alpha})/(\mathrm{GeV})^2,\\
\Omega_2&=&  (p_D^{\mu} + p_C^{\mu}) (p_D^{\nu} + p_C^{\nu})T_{\mu\nu\alpha\beta}(p_B^{\alpha} - p_A^{\alpha}) (p_B^{\beta} - p_A^{\beta})/(\mathrm{GeV})^4,
\end{eqnarray}
\end{widetext}
where $T_{\mu\alpha} \equiv -g_{\mu\alpha} +  p_{\mu} p_{\alpha}/m_R^2$, and $T_{\mu\nu \alpha\beta} = (1/2) \left(T_{\mu\alpha} T_{\nu\beta} + T_{\mu \beta} T_{\nu\alpha}\right)- (1/3) T_{\mu\nu} T_{\alpha\beta}.$ The sign of the spin--1 factor depends on the ordering of the
resonance's decay products. The ordering choice for the
different spin factors is given in
Table~\ref{tab:spinFactorOrdering}. Care is taken to ensure that $CP$--conjugate
amplitudes such as $D \to K^{*}K_S^0$ and $\bar{D} \to \bar{K}^* K_S^0$ are assigned the same spin factor (with the same sign). The Blatt--Weisskopf penetration factors, $F_R $ and $F_D$ are described in ~\cite{blattweiss} and are listed in Table~\ref{tab:penetration}. They rely on one free parameter $\mathcal{R}$, the meson radius and on the momentum of the decay particles in the rest frame of the parent. For consistency with other experiments~\cite{e687} $\mathcal{R}$ is fixed to be 5 (GeV/$c$)$^{-1}$, and 1.5 (GeV/$c$)$^{-1}$ for the $D^0$ meson and intermediate resonances, respectively.
\begin{table}
\begin{center}
\caption{Particle ordering in spin~1 resonances.\label{tab:spinFactorOrdering}}
\begin{tabular}{lccc}
\hline\hline
Decay & A & B & C\\
\hline
$D^0 \to K^{*0}K_S^0, \;K^{*0}\to K^- \pi^+$ & $K^-$ & $\pi^+$ & $K_S^0$\\
$D^0 \to K^{*+}K^-, \;K^{*+}\to K^0_S \pi^+$ & $K_S^0$ & $\pi^+$ & $K^-$\\
$D^0 \to \rho^-\pi^+, \;\rho^-\to K^- K_S^0$ & $K_S^0$ & $K^-$ & $\pi^+$
\\\hline\hline
\end{tabular}
\end{center}
\end{table}

\begin{table}
\begin{center}
\caption{The Blatt--Weisskopf penetration factors for the mesons of different spin. The momentum of either daughter in the meson rest frame, $p_r$ and the momentum of either daughter in the candidate rest frame $p_{AB}$, differ due to the use of the mass of the meson or the invariant mass from the two--daughter combination.\label{tab:penetration}}
\begin{tabular}{lc}
\hline \hline
Spin & $F_R$\\
\hline
0 & 1\\
1 & $\sqrt{\frac{1+ \mathcal{R}^2 p_r^2}{1+ \mathcal{R}^2 p_{AB}^2}}$\\
2 & $\sqrt{\frac{9 + 3\mathcal{R}^2 p_r^2 + \mathcal{R}^4 p_r^4}{9 + 3\mathcal{R}^2 p_{AB}^2 + \mathcal{R}^4 p_{AB}^4}}$\\

\hline \hline
\end{tabular}
\end{center}
\end{table}

The signal model is made from a sum of intermediate resonances so that the total matrix element is given by 
\begin{equation}
\mathcal{M}_{\mathrm{tot}} = \sum_j a_j e^{i\phi_j} \mathcal{M}_j,
\end{equation}
where $M_j$ is as defined in ~\equationref{eq:fullmi}, and $a_j$ and $\phi_j$ are the relative amplitudes and phases between different resonances.
There are a number of intermediate resonances which could contribute to the decays \dtofav and \dtosup. The presence of the $K^{*}(892)^\pm$ resonance is clearly visible in the Dalitz plot distributions for both the favored and suppressed decays. For this resonance $a_j$ is set to one and $\phi_j$ to zero, as all other amplitudes and phases can be measured relative to this component. The Breit--Wigner lineshape is used for this resonance.  The presence of the $K^{*}_0(1430)^\pm$ is also required. It is described with the LASS lineshape~\cite{lass} with the modifications described in~\cite{babar}. This parametrization is used as it includes an S--wave contribution and can account for the broad shape of any non--resonant component. 
The finite sample size limits the number of contributions that can be fitted to the data, without a large number of components having fit fractions consistent with zero, where the fit fraction $F_i$ for resonance $i$ is defined as
\begin{equation}
F_i=\frac{\int dm^2_{K_S^0K}dm^2_{K\pi}|\mathcal{M}_i(m^2_{K_S^0K},m^2_{K\pi})|^2}{\int dm^2_{K_S^0K}dm^2_{K\pi}|\sum^n_{k=1} \mathcal{M}_k(m^2_{K_S^0K},m^2_{K\pi})|^2}.
\end{equation}
 The number of intermediate resonances fit to the favored decay channel is six as this usually produces a fit where no more than two resonances have fit fractions consistent with zero. The number of intermediate resonances fit to the suppressed decay channel is five, due to the lower signal yield of this decay. Hence in addition to the $K^{*}(892)^\pm$ and $K^{*}_{0}(1430)^\pm$ a further four (three) resonances are chosen to make the favored (suppressed) model. These resonances are taken from all known possibilities that decay into either $K_S^0\pi$, $K^\pm\pi^\mp$, or $K_S^0K^\pm$.  These are the $a_0(980,1450)^+,$  $a_2(1320)^+,$ $K^{*}(892,1410,1680)^{0/\pm}$, $K^{*}_0(1430)^{0/\pm}$, $K^{*}_2(1430)^{0/\pm}$, and $\rho(1450,1700)^+$, where for each resonance the mass and width are taken from the PDG~\cite{PDG}. The Breit--Wigner lineshape is used for all these resonances except for the  $K^{*}_0(1430)^\pm$ which uses the lineshape in~\cite{babar}, and $a_0(980)$ where the Flatt\'e~\cite{flatte} lineshape is used.

\subsection{Detector efficiency and background modeling}
\label{sec:ampeffmodel}

The detector efficiency, which varies over the Dalitz plot, and the distributions of the background events need to be taken into account when fitting the data to extract the amplitudes and phases of each intermediate resonance.
The models for the detector efficiency are determined from Monte Carlo simulation. The efficiency shapes for \cleoiii\ and CLEO-c data differ due to the different $D^0$ momentum distributions and detector configurations. The simulated events are generated non--resonantly and hence are distributed evenly over the Dalitz plot. The shape of the relative efficiency is largely flat although a reduction in efficiency is seen in regions where one of the particles has low momentum. The simulation samples for CLEO-c are large. The probability density function (PDF) can be integrated using the simulated events, and as they are distributed according to the efficiency shape there is no need to parametrize the CLEO-c efficiency. The Monte Carlo statistics available for \cleoiii\ are not as large and in this case the efficiency shape is parametrized by a second--order polynomial in $(m^2_{K_S^0K},m^2_{K\pi})$.

The flavor--tagged data samples have a purity of 90$\%$ or greater as described in ~\secref{sec:datacleo3} and ~\secref{sec:datacleoc}. For \cleoiii\ data the background shape is determined from three sidebands in data. Two of the sidebands require $\Delta m_D$ to be less than 1 \mevc$^2$  from  its nominal value and either $1.7<m_{D^0}<1.8$ \gevcc\ or $1.9<m_{D^0}<1.93$ \gevcc. The second sideband has a more restricted mass range in order to avoid contamination from $D^0 \to \pi^+\pi^-\pi^+\pi^-$ events that peak at higher reconstructed $D^0$ mass. The third sideband requires $m_{D^0}$ to be within 15 \mevc$^2$ of its nominal value and 0.148 $<\Delta m_D<0.155$ \gevcc. The candidate events from all three sidebands are combined as there are insufficient data for separate fits. The parametrization of sideband data over the Dalitz plot is determined empirically. It is modeled as a sum of four incoherent pseudo--resonances where the masses, widths and spins of the pseudo--resonances are adjusted until a good fit is achieved. Maximum likelihood fits are performed on background samples to determine the background shapes. The free parameters in these fits are the relative contributions of the pseudo--resonances.  The background shapes are determined separately for the favored and suppressed decays as there are different contributions from misreconstructed $D$ meson decays. For CLEO-c data the background shape is determined from the combined Monte Carlo samples described in ~\secref{sec:datacleoc} in the same way as the \cleoiii\ samples. The data sidebands are not used due to insufficient yields to describe the shape on the Dalitz plot.

\subsection{Dalitz plot probability density function}
The data are fit using a similar method to previous CLEO analyses~\cite{pene}. In order to describe the event density distribution on the
Dalitz plot a probability density function is used. For the isobar model fit the PDF is given by
\begin{equation}
\begin{split}
 \mathcal{P}_j&(m^2_{K_S^0K},m^2_{K\pi}) =\\&
 f_{\mathrm{sig},j}|\mathcal{M}(m^2_{K_S^0K},m^2_{K\pi})|^2\varepsilon_j(m^2_{K_S^0K},m^2_{K\pi}) +\\
& (1-f_{\mathrm{sig},j})B_j(m^2_{K_S^0K},m^2_{K\pi}),
\end{split}
\end{equation}
where $\varepsilon(m^2_{K_S^0K},m^2_{K\pi})$ and $B(m^2_{K_S^0K},m^2_{K\pi})$ represent the functions that describe the efficiency and background shapes, respectively, across the Dalitz plot, $f_{\mathrm{sig}}$ is the signal fraction and $j$ denotes either the CLEO~III or CLEO-c data. The matrix element $\mathcal{M}$ is defined in \secref{sec:ampdefsdet}. The PDF components $\varepsilon$, $|M|^2$, and $B$ are all normalized individually over the Dalitz plot, which leads to $\mathcal{P}$ being normalized to unity as well. 
In the maximum likelihood fit to determine the isobar model the only free parameters are the real and imaginary part of each contributing intermediate resonance except the $K^*(892)^{^\pm}$ where the amplitude and phase are fixed to one and zero, respectively. The background shapes and \cleoiii\ efficiency shapes are fixed from fits to the individual relevant samples described in \secref{sec:ampeffmodel}, and the CLEO-c efficiency is taken from the large Monte Carlo sample as described in \secref{sec:ampeffmodel}. The total likelihood to determine the isobar model is given by 
\begin{widetext}
\begin{equation}
\label{eq:PDF}
\mathcal{L} =\prod_{\parbox{4.0em}{\scriptsize\center \mbox{}\vspace{-1ex}\\j=CLEO-c, \newline \cleoiii}}\left[ \prod_i f_{\mathrm{sig},j}|\mathcal{M}(m^2_{K_S^0K, i},m^2_{K\pi, i})|^2\varepsilon_j(m^2_{K_S^0K,i},m^2_{K\pi,i}) + (1-f_{\mathrm{sig},j})B_j(m^2_{K_S^0K,i},m^2_{K\pi,i}) \right],
\end{equation}
\end{widetext}
where $i$ labels each event, in data sample $j$.

Using the efficiency and background model PDFs, the likelihood given in ~\equationref{eq:PDF} is constructed where the matrix element contains a combination of resonances. All possible isobar models, constructed from the full set of considered resonances, that have the allowed total number of components, and include the $K^*(892)^\pm$ and $K^*_0(1430)^\pm$ contributions, are tested when fitting to the data.

The contributions of the resonances in the $K^\pm\pi^\mp$ channel are expected to be suppressed in comparison to resonances in the $K_S^0\pi$ or $K_S^0K$ channel as this process proceeds by a suppressed $W$ boson exchange. Hence models where there are large contributions in the $K^\mp\pi^\pm$ channel are considered unphysical and are discarded. Also discarded are models where the total fit fraction is greater than 150$\%$ as these fits usually contain contributions that are highly correlated with each other and contribute very little independently to the fit.

To estimate the quality of the fit a binned $\chi^2$ quantity is computed. The Dalitz plot is split into square bins of size 0.18 GeV$^2/c^4$ vs. 0.18 GeV$^2/c^4$. There are a total of 67 bins which contain at least part of the allowed phase space. Bins that contain less than 10 events are merged with the neighboring bins until the content is greater than 10.  The $\chi^2$ is defined as $(N_{\mathrm{obs}}-N_{\mathrm{exp}})^2/N_{\mathrm{exp}}$, where $N_{\mathrm{obs}}$ and $N_{\mathrm{exp}}$ are the number of observed and expected events, respectively. For a given fit, the $\chi^2$ per degree of freedom (dof) is evaluated, and used to discriminate between different fit models.

Models that contain the $K^{*}(892)^0$ resonance are interesting as they can be used to test SU(3)--flavor symmetry predictions~\cite{jon}. As the best--fit models in both the favored and suppressed case do not contain this contribution, an additional model is constructed for each decay with this resonance required to be present. The models with the $K^*(892)^0$ resonance are labeled as model 1 and the best--fit models are labeled as model 2. Table~\ref{tab:modsum} summarizes the resonance contribution, the $\chi^2$/dof, and the figures and tables that display the results for each model.

\begin{table*}[htbp]

\begin{center}
\caption{A summary of the resonance contributions and $\chi^2$/dof for each model. The figures showing the projection and the table listing the fit information are also given.\label{tab:modsum}}
\begin{tabular}{llcll}
\hline \hline
Model & Resonance & $\chi^2$/dof & Projection & Results \\ \hline
Favored 1 &  $K^*(892)^+$, $K^*(892)^0$, $K^*_0(1430)^+$,  $K^*(1410)^0$,  $K^*(1680)^+$, $a_0(1450)^-$ & 42.2/26 & ~\figref{fig:modelproj1} & ~\tabref{tab:ampfavres} \\
Favored 2 &  $K^*(892)^+$, $\rho(1700)^-$, $K^*_0(1430)^+$,  $K^*(1410)^0$,  $K^*(1680)^+$, $a_0(1450)^-$ & 37.3/26 & ~\figref{fig:modelproj2} & ~\tabref{tab:ampfavres2} \\

Suppressed 1 &  $K^*(892)^-$, $K^*(892)^0$, $K^*_0(1430)^-$, $a_0(1450)^+$, $\rho(1700)^+$  & 50.9/30 & ~\figref{fig:modelproj3} & ~\tabref{tab:ampsupres} \\
Suppressed 2 &  $K^*(892)^-$, $K^*_2(1430)^0$, $K^*_0(1430)^-$, $a_0(1450)^+$, $\rho(1700)^+$  & 46.5/30 & ~\figref{fig:modelproj4} & ~\tabref{tab:ampsupres2} \\

\hline \hline
\end{tabular}
\end{center}

\end{table*}

 \subsection{Systematic uncertainties}
 The systematic uncertainties for the measured amplitudes and phases are determined by performing fits with the same resonances as the chosen models in an alternate manner. The shift between the alternate and default fit result is taken to be the systematic uncertainty on that parameter. The relative sizes of the systematic uncertainties are discussed in relation to the statistical uncertainty on the fitted amplitudes and phases and the fit fraction.

The leading systematic uncertainty arises from the choice of values for the meson radii in the Blatt--Weiskopff penetration factors. In the alternate fit these are changed to 1.5 (GeV/$c$)$^{-1}$ and 0.5 (GeV/$c$)$^{-1}$ for the $D^0$ meson and intermediate resonances, respectively. It is observed that the changes in some of the fitted parameters in the favored models are up to three times the size of the statistical uncertainty, whereas in the suppressed models the changes are of the same order as the statistical uncertainty. This is understood as the favored models have a larger contribution from resonances where the pole mass lies outside the phase space. The change in the value of the penetration factor increases with increasing distance from the pole mass, and hence this effect is a larger source of systematic uncertainty for the favored model.

 The other leading systematic uncertainty is attributed to uncertainties in the measured masses and widths of the resonances, and the parameters of the $K^{*}_0(1430)^\pm$ lineshape. The masses and widths of the resonances, and the parameters of the $K^{*}_0(1430)^\pm$ lineshape are varied, each in turn by one standard deviation from their measured values. The shifts on the fitted amplitudes and phases are added in quadrature for each alteration. Overall the total systematic uncertainty arising from these changes is of order half the statistical uncertainty. 

The PDF used to arrive at the default models neglects the DCS amplitude that is present in the flavor tag decays. Although the DCS rate is much smaller than the CF rate, the interference between the two amplitudes is a significant effect in the quantum--correlated CLEO-c data. To determine the size of this effect the data are fit with a modified PDF for the CLEO-c data. In the case of the favored data, the alternate PDF uses a total amplitude given by $\mathcal{A}_{\mathrm{tot}} = \mathcal{A}_{\mathrm{fav}} + 0.05 \times 0.7 \times \mathcal{A}_{\mathrm{sup}} e^{i\delta}$. The parameters of the suppressed model are fixed to their default values, while the parameters of the favored model are free. The factor 0.05 arises from the approximate ratio of DCS to CF amplitudes in neutral $D$ meson decays to $K\pi$. There is, however, dilution as the $K\pi\pi\pi$ and $K\pi\pi^0$ tags are not fully coherent. The factor of 0.7 accounts for the weighted average of the fully coherent $K\pi$ decay and the coherence factors for the other two tags measured in~\cite{jim}. Additionally there is an unknown absolute phase $\delta$ between the favored and suppressed amplitude models. To account for this, the fit is repeated assuming a phase difference of zero, $\pi/2$, $\pi$, and $3\pi/2$. The largest shifts in the fitted parameters are observed for a phase difference of zero and these are taken to be the systematic uncertainty. In general the systematic uncertainty due to this effect is about 25$\%$ of the statistical error.
In the \cleoiii\ data there is no quantum--correlation, however mistags occur when a track of the wrong charge is misidentified as the pion from the $D^{*}$. The mistag rate is determined by measuring the relative rate of, for example, $D^0\to K^+\pi^-\pi^+\pi-$ and $D^0->K^-\pi^+\pi^-\pi^+$ candidates associated with a $D^{*+} \to  D^0 \pi^+$ decay chain, and attributing to mistags the excess of doubly--Cabibbo--suppressed candidates that is observed above expectation.  The mistag rate is found to be around 0.6$\%$. The systematic uncertainty due to this is determined to be negligible in comparison to other possible sources of bias.

 The systematic uncertainty due to the background fraction is evaluated by modifying the assigned value of the purity in the fit PDF by the uncertainty on its measured value. In addition, alternate background models are determined using a fit to the background samples with fewer pseudo--resonances. The alternate parametrizations are used in the fit to signal data, and the observed shifts are used as the systematic uncertainty. The uncertainty on the efficiency shape over the Dalitz plot is assessed using alternate parametrizations for the efficiency shape. In the case of \cleoiii\ a third--order polynomial is used instead of the default second--order polynomial. In the case of CLEO-c the efficiency is parametrized by a fourth--order polynomial instead of using the Monte--Carlo--simulated sample directly. The uncertainties on efficiency, purity, and background shapes each contribute an uncertainty of approximately $10\%$ of the statistical uncertainty. The default fit does not take into account the detector resolution on the Dalitz plot, and hence all data points are slightly shifted from their true positions. To account for the resulting uncertainty, the $K_S^0$, $K$, and $\pi$ momenta are smeared in data, according to parameters determined from Monte Carlo simulation, and the fit is performed on the smeared sample. This systematic uncertainty is negligible in comparison to other contributions. The validity of the fitter is assessed via fits to many samples of simulated data generated from the PDFs. A small uncertainty of order 10$\%$ of the statistical uncertainty is assigned due to small residual biases between the generated and fitted parameters.

A summary of the systematic uncertainties for each effect as a fraction of the statistical uncertainty on the amplitudes, phases, and fit fractions for models 1 are given in Table~\ref{tab:systsumfav} and~\ref{tab:systsumsup}. The relative sizes of the systematic uncertainties are similar for models 2.

\begin{table*}[htbp]
\caption{Systematic uncertainties on $F_i$, $|a_i|$, and $\phi_i$ in units of statistical standard deviations ($\sigma$) for the decay \dtofav. The different contributions are: (I) Blatt--Weisskopf penetration factors; (II) mass and width of resonances; (III) quantum correlations; (IV) CLEO III mistag; (V) background models; (VI) sample purities; (VII) acceptance; (VIII) resolution; (IX) fitter bias.\label{tab:systsumfav}}
\begin{tabular}[c]{lcccccccccc}
\hline\hline
 & \multicolumn{9}{c}{Source ($\sigma$)} & Total ($\sigma$)\\
 & I & II & III & IV & V & VI & VII & VIII & IX & \\
\hline
$K^*(892)^+K^-$  $F_i$ & 0.36 & 0.33 & 0.13 & 0.08 & 0.08 & 0.15 & 0.16 & 0.05 & 0.18 & 0.59\\
$K^*(892)^0K_S^0$  $|a_i|$ & 0.19 & 0.24 & 0.06 & 0.09 & 0.12 & 0.16 & 0.14 & 0.06 & 0.06 & 0.42\\
$K^*(892)^0K_S^0$  $\phi_i$ & 0.87 & 0.28 & 0.11 & 0.02 & 0.10 & 0.08 & 0.11 & 0.07 & 0.05 & 0.94\\
$K^*(892)^0K_S^0$  $F_i$ & 0.21 & 0.21 & 0.05 & 0.10 & 0.13 & 0.19 & 0.16 & 0.05 & 0.21 & 0.47\\
$K^*(1410)^0K_S^0$  $|a_i|$ & 4.35 & 0.85 & 0.23 & 0.04 & 0.09 & 0.07 & 0.09 & 0.01 & 0.12 & 4.44\\
$K^*(1410)^0K_S^0$  $\phi_i$ & 1.13 & 0.39 & 0.05 & 0.07 & 0.17 & 0.13 & 0.12 & 0.01 & 0.13 & 1.23\\
$K^*(1410)^0K_S^0$  $F_i$ & 0.96 & 0.35 & 0.32 & 0.04 & 0.07 & 0.07 & 0.06 & 0.02 & 0.05 & 1.08\\
$K^*(1680)^+K^-$  $|a_i|$ & 1.53 & 0.41 & 0.07 & 0.08 & 0.10 & 0.15 & 0.14 & 0.03 & 0.21 & 1.62\\
$K^*(1680)^+K^-$  $\phi_i$ & 1.15 & 0.41 & 0.03 & 0.09 & 0.11 & 0.17 & 0.16 & 0.02 & 0.26 & 1.28\\
$K^*(1680)^+K^-$  $F_i$ & 1.20 & 0.32 & 0.17 & 0.00 & 0.03 & 0.02 & 0.12 & 0.02 & 0.09 & 1.26\\
$K_0^*(1430)^+K^-$  $|a_i|$ & 1.44 & 0.32 & 0.20 & 0.08 & 0.14 & 0.15 & 0.21 & 0.02 & 0.39 & 1.57\\
$K_0^*(1430)^+K^-$  $\phi_i$ & 0.49 & 0.19 & 0.08 & 0.02 & 0.08 & 0.04 & 0.13 & 0.00 & 0.05 & 0.56\\
$K_0^*(1430)^+K^-$  $F_i$ & 1.51 & 0.35 & 0.15 & 0.07 & 0.12 & 0.13 & 0.18 & 0.03 & 0.16 & 1.59\\
$a_0(1450)^-\pi^+$  $|a_i|$ & 0.41 & 0.35 & 0.20 & 0.04 & 0.08 & 0.08 & 0.15 & 0.02 & 0.08 & 0.61\\
$a_0(1450)^-\pi^+$  $\phi_i$ & 0.42 & 0.44 & 0.18 & 0.03 & 0.04 & 0.06 & 0.08 & 0.04 & 0.06 & 0.64\\
$a_0(1450)^-\pi^+$  $F_i$ & 0.54 & 0.24 & 0.29 & 0.02 & 0.10 & 0.04 & 0.22 & 0.01 & 0.73 & 1.01\\
\hline\hline
\end{tabular}
\end{table*}

\begin{table*}[htbp]
\caption{Systematic uncertainties on $F_i$, $|a_i|$, and $\phi_i$ in units of statistical standard deviations ($\sigma$) for the decay \dtosup. The different contributions are: (I) Blatt--Weisskopf penetration factors; (II) mass and width of resonances; (III) quantum correlations; (IV) CLEO III mistag; (V) background models; (VI) sample purities; (VII) acceptance; (VIII) resolution; (IX) fitter bias.\label{tab:systsumsup}}
\begin{tabular}[c]{lcccccccccc}
\hline\hline
 & \multicolumn{9}{c}{Source ($\sigma$)} & Total ($\sigma$)\\
 & I & II & III & IV & V & VI & VII & VIII & IX & \\
\hline
$K^*(892)^-K^+$  $F_i$ & 0.23 & 0.21 & 0.20 & 0.05 & 0.03 & 0.04 & 0.11 & 0.00 & 0.02 & 0.39\\
$K^*(892)^0K_S^0$  $|a_i|$ & 0.13 & 0.37 & 0.04 & 0.03 & 0.03 & 0.02 & 0.05 & 0.00 & 0.08 & 0.41\\
$K^*(892)^0K_S^0$  $\phi_i$ & 0.20 & 0.24 & 0.03 & 0.01 & 0.04 & 0.02 & 0.12 & 0.00 & 0.04 & 0.34\\
$K^*(892)^0K_S^0$  $F_i$ & 0.10 & 0.23 & 0.01 & 0.02 & 0.02 & 0.04 & 0.07 & 0.01 & 0.07 & 0.27\\
$K_0^*(1430)^-K^+$  $|a_i|$ & 0.33 & 0.33 & 0.10 & 0.03 & 0.02 & 0.03 & 0.08 & 0.00 & 0.19 & 0.52\\
$K_0^*(1430)^-K^+$  $\phi_i$ & 0.23 & 0.43 & 0.07 & 0.02 & 0.01 & 0.02 & 0.03 & 0.00 & 0.13 & 0.51\\
$K_0^*(1430)^-K^+$  $F_i$ & 0.31 & 0.37 & 0.09 & 0.03 & 0.02 & 0.01 & 0.08 & 0.00 & 0.15 & 0.52\\
$a_0(1450)^+\pi^-$  $|a_i|$ & 0.58 & 0.56 & 0.08 & 0.02 & 0.01 & 0.02 & 0.04 & 0.01 & 0.06 & 0.81\\
$a_0(1450)^+\pi^-$  $\phi_i$ & 0.46 & 0.58 & 0.06 & 0.01 & 0.02 & 0.01 & 0.04 & 0.00 & 0.04 & 0.75\\
$a_0(1450)^+\pi^-$  $F_i$ & 0.43 & 0.24 & 0.03 & 0.00 & 0.01 & 0.01 & 0.03 & 0.01 & 0.17 & 0.52\\
$\rho (1700)^+\pi^-$  $|a_i|$ & 1.25 & 1.22 & 0.05 & 0.00 & 0.02 & 0.01 & 0.07 & 0.01 & 0.09 & 1.75\\
$\rho (1700)^+\pi^-$  $\phi_i$ & 0.36 & 0.71 & 0.09 & 0.00 & 0.01 & 0.01 & 0.06 & 0.00 & 0.02 & 0.80\\
$\rho (1700)^+\pi^-$  $F_i$ & 1.01 & 0.68 & 0.01 & 0.01 & 0.01 & 0.01 & 0.09 & 0.01 & 0.02 & 1.22\\
\hline\hline
\end{tabular}
\end{table*}

\subsection{Results of the amplitude studies}
\label{sec:ampresults}
\begin{figure*}[htbp]

\begin{center}
   \includegraphics[width=0.9\textwidth]{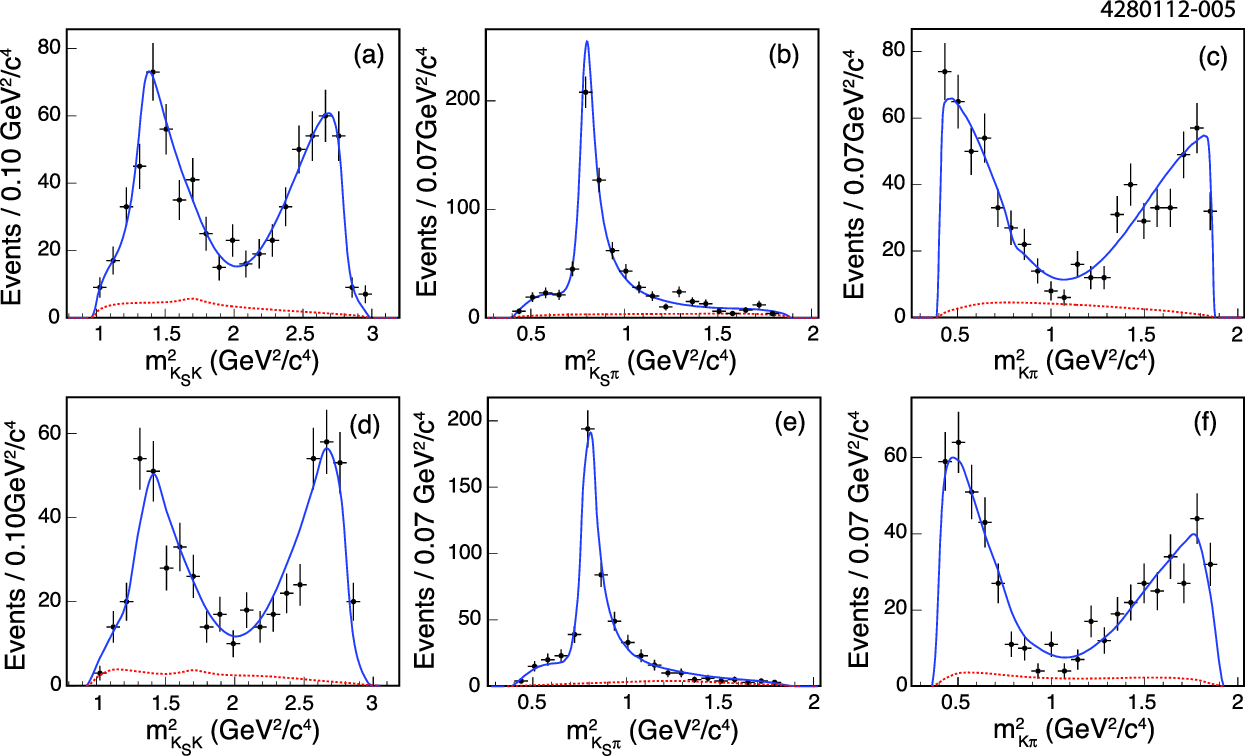}
\end{center}

\caption{Fit projections for the \dtofav\ decay mode using favored model 1, the best model containing the $K^*(892)^0$ resonance. The data are shown as points and the total fit projection is overlaid with the solid line. The projection with the dotted line is the contribution of background. Plots (a)--(c)((d)--(f)) show the \cleoiii\ (CLEO-c) data for the three pairs of invariant mass distributions.
 \label{fig:modelproj1}}
\end{figure*}

\begin{figure*}[htbp]

\begin{center}
   \includegraphics[width=0.9\textwidth]{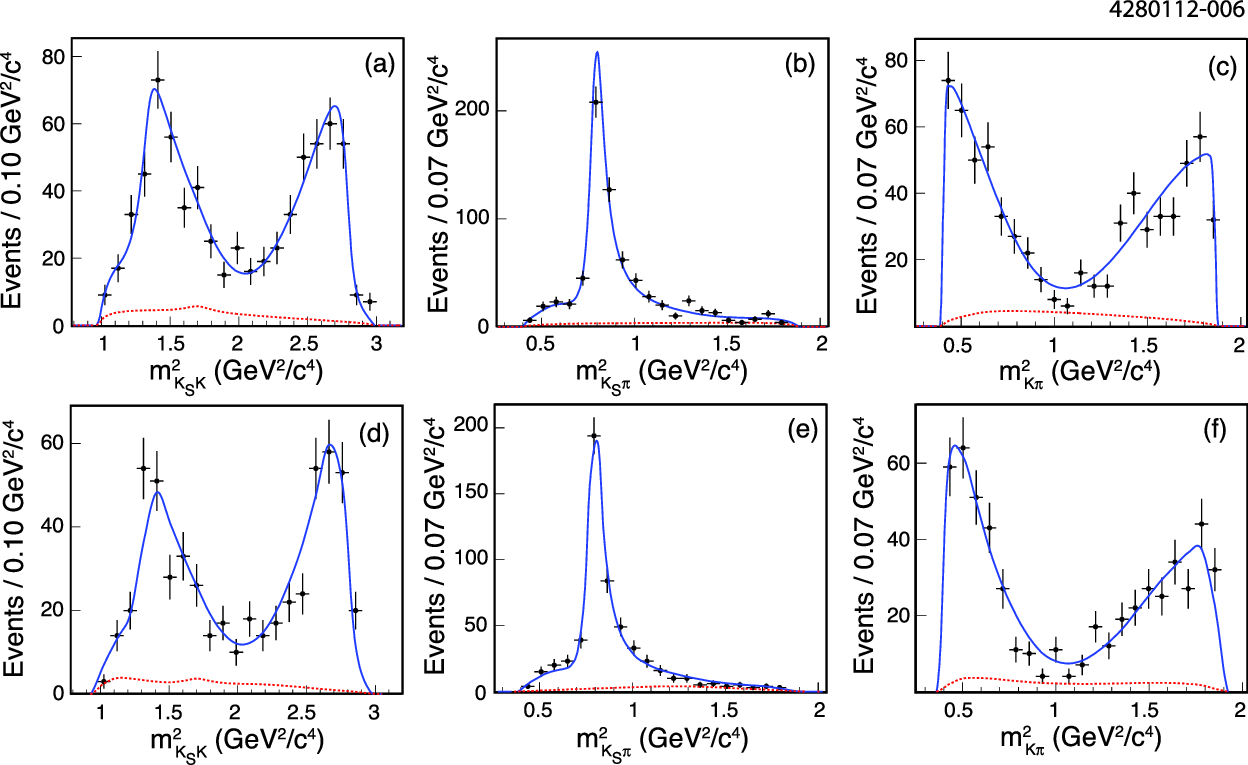}
\end{center}
\caption{Fit projections for the \dtofav\ decay mode using favored model 2, the best model. The data are shown as points and the total fit projection is overlaid with the solid line. The projection with the dotted line is the contribution of background. Plots (a)--(c)((d)--(f)) show the \cleoiii\ (CLEO-c) data for the three pairs of invariant mass distributions. \label{fig:modelproj2}}
\end{figure*}

\begin{figure*}[htbp]
\begin{center}
   \includegraphics[width=0.9\textwidth]{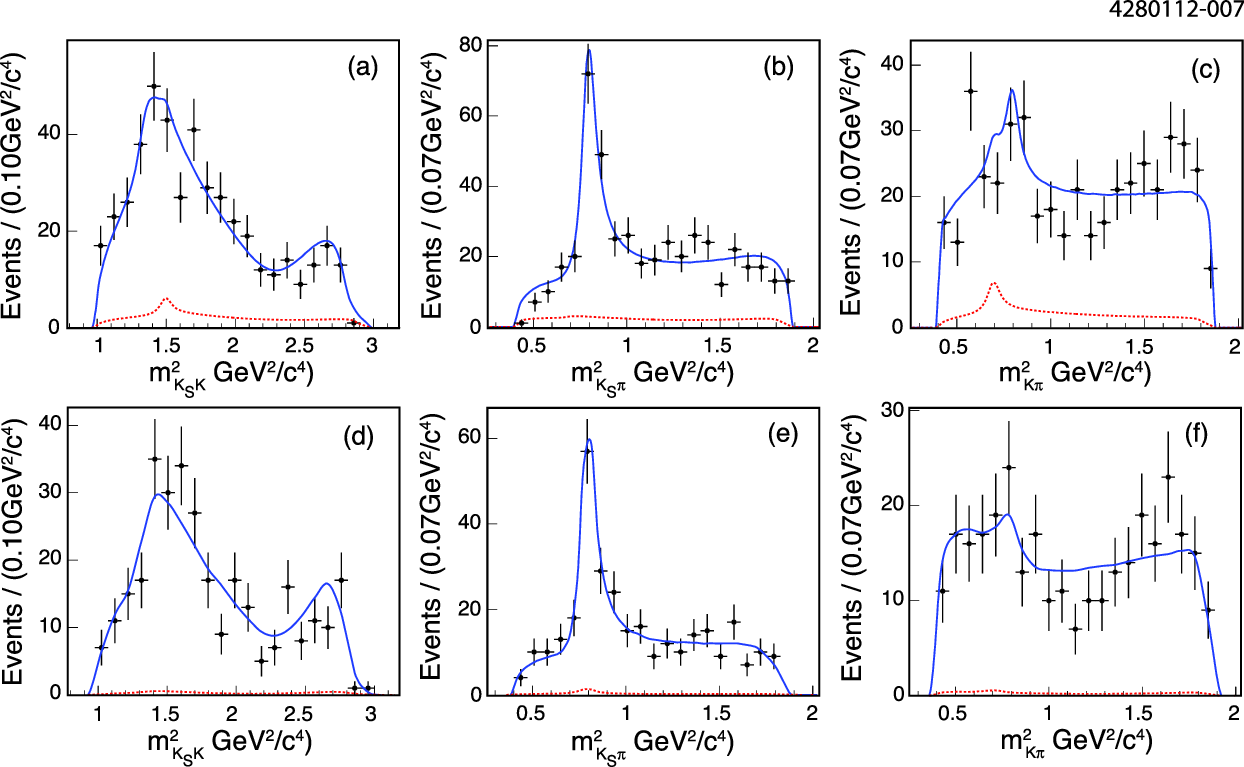}
\end{center}
\caption{Fit projections for the \dtosup\ decay mode using suppressed model 1, the best model containing the $K^*(892)^0$ resonance. The data are shown as points and the total fit projection is overlaid with the solid line. The projection with the dotted line is the contribution of background. Plots (a)--(c)((d)--(f)) show the \cleoiii\ (CLEO-c) data for the three pairs of invariant mass distributions.\label{fig:modelproj3}}
\end{figure*}

\begin{figure*}[htbp]
\begin{center}
    \includegraphics[width=0.9\textwidth]{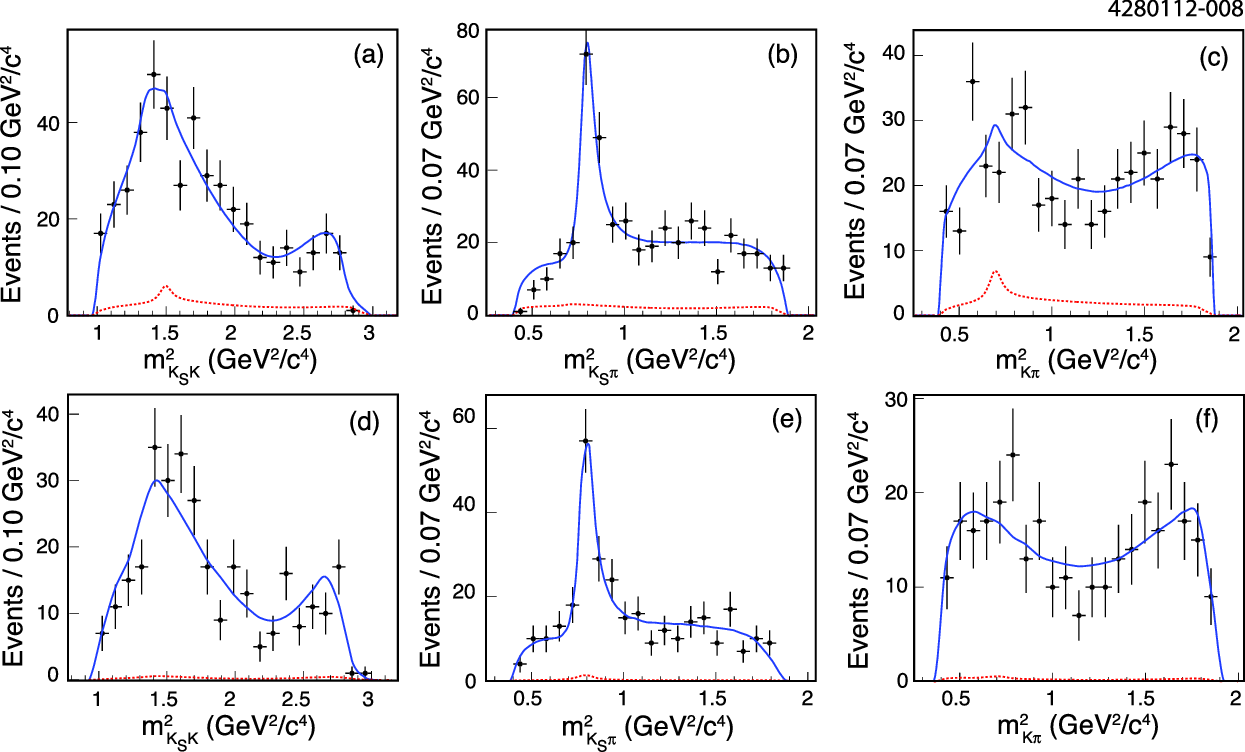}
  \end{center}
\caption{Fit projections for the \dtosup\ decay mode using suppressed model 2, the best model. The top plots show the projection of the fit on the \cleoiii\ data while the lower plots show the projection onto CLEO-c data. The data are shown as points and the total fit projection is overlaid with the solid line. The projection with the dotted line is the contribution of background. Plots (a)--(c)((d)--(f)) show the \cleoiii\ (CLEO-c) data for the three pairs of invariant mass distributions.  \label{fig:modelproj4}}
\end{figure*}

The results of the fit are listed in Tables ~\ref{tab:ampfavres}--\ref{tab:ampsupres2}. These tables show the amplitude, phase and fit fraction for each contributing resonance, and the statistical and systematic uncertainty for each of these parameters. The fit projections for all these models are given in ~Figs.~\ref{fig:modelproj1}--\ref{fig:modelproj4}.

The fit projections show good agreement with the data. The difference in fit projections between model 1 and and model 2 in either of the decay modes is very small. The $K^*(892)^\pm$ component, clearly visible in the data, is seen to contribute significantly to all four models. This large, common resonant structure gives the expectation of significant coherence between the two decay modes, particularly in the region of the $K^*(892)^\pm$ resonance. The data for the suppressed decay show a broad extended shape either side of the $K^*(892)^\pm$ resonance in the $m^2_{K_S^0\pi}$ projection which is not present in the favored decay.
This difference is apparent in the models as the favored models have a very small contribution from the  $K^*(1430)^0$ resonance which is described by a broad S--wave lineshape, while in the suppressed models this is a significant contribution to the overall model.
In the favored models the fit fraction of the $K^*(1410)^0$ resonance is large, which is unexpected as the $K^\mp\pi^\pm$ channel proceeds by a suppressed $W$ boson exchange. However the large fit fraction can be understood by considering the destructive interference between $K^*(1410)^0$ and $K^*(1680)^\pm$ which is also significant. This interference between two resonances can be quantified by calculating the interference fit fraction between any two resonances $(j,l)$, and is given by
\begin{equation}
\label{eq:interference}
  I_{j,l}=\frac{Re{\int dm^2_{K_S^0K}dm^2_{K\pi}a_je^{i\phi_j}\mathcal{M}_ja^*_le^{-i\phi_l}\mathcal{M}_l}}{\int dm^2_{K_S^0K}dm^2_{K\pi}|\sum^n_{k=1} \mathcal{M}_k(m^2_{K_S^0K},m^2_{K\pi})|^2}.
\end{equation}
 For these two resonances the interference fit fraction is $-21\%$ ($-27\%$) for model 1 (2). The full tables of interference fit fractions are provided in Appendix A. 
There is similarity between models 1 and 2 for each decay mode, as the contribution of the $K^*(892)^0$ is small. From SU(3) flavor symmetry predictions~\cite{jon}, the expected ratio of amplitudes between $D^0\to \bar{K^{*0}}K^0$ and $D^0\to K^{*+}K^-$ is 0.138$\pm$0.033. Using favored model 1 we observe a ratio of 0.2 $\pm$ 0.1. Although this is consistent with the prediction, it should be noted that the fit fraction of $K^*(892)^0$ is small and consistent with zero at the one standard deviation level. An isobar model determined with a larger sample than analyzed here will be required to test the prediction with more meaningful precision.

\begin{table*}[htbp]
\caption{The results of the Dalitz plot fit for the decay \dtofav. This is the best model including the neutral $K^*(892)^0$ resonance, favored model 1. The amplitudes, phases, and fit fractions are given. The first uncertainty is statistical, and the second systematic.\label{tab:ampfavres}}
\begin{tabular}[c]{lr@{$\pm$}c@{$\pm$}rr@{$\pm$}c@{$\pm$}rr@{$\pm$}c@{$\pm$}r}

\hline\hline
Component & \multicolumn{3}{c}{Amplitude} & \multicolumn{3}{c}{Phase ($^\circ$)} & \multicolumn{3}{c}{Fraction (\%)}\\
\hline

$D^0\to K^*(892)^+K^-$ & \multicolumn{3}{c}{1.0 (fixed)} & \multicolumn{3}{c}{0.0 (fixed)} & 67.6 & 6.4 & 3.8 \\
$D^0\to K^*(892)^0K_S^0$ & 0.2 & 0.1 & 0.0 & 96.1 & 13.2 & 12.4 & 1.8 & 1.7 & 0.8 \\
$D^0\to K^*(1410)^0K_S^0$ & 8.4 & 0.7 & 3.0 & $-16.7$ & 7.5 & 9.2 & 23.2 & 3.9 & 4.2 \\
$D^0\to K^*(1680)^+K^-$ & 15.3 & 2.3 & 3.8 & 148.3 & 11.1 & 14.2 & 22.7 & 4.4 & 5.6 \\
$D^0\to K_0^*(1430)^+K^-$ & 1.3 & 0.3 & 0.5 & $-101.0$ & 22.0 & 12.3 & 2.5 & 1.2 & 2.0 \\
$D^0\to a_0(1450)^-\pi^+$ & 0.4 & 0.2 & 0.1 & $-26.8$ & 42.3 & 27.2 & 1.0 & 0.9 & 0.9 \\

\hline\hline\end{tabular}
\end{table*}
\begin{table*}[htbp]
\caption{The results of the Dalitz plot fit for the decay \dtofav. This is the model with the lowest $\chi^2$ per degree of freedom, favored model 2. The amplitudes, phases, and fit fractions are given. The first uncertainty is statistical, and the second systematic.\label{tab:ampfavres2}}
\begin{tabular}[c]{lr@{$\pm$}c@{$\pm$}rr@{$\pm$}c@{$\pm$}rr@{$\pm$}c@{$\pm$}r}

\hline\hline
Component & \multicolumn{3}{c}{Amplitude} & \multicolumn{3}{c}{Phase ($^\circ$)} & \multicolumn{3}{c}{Fraction (\%)}\\
\hline

$D^0\to K^*(892)^+K^-$ & \multicolumn{3}{c}{1.0 (fixed)} & \multicolumn{3}{c}{0.0 (fixed)} & 62.2 & 4.8 & 3.9 \\
$D^0\to K^*(1410)^0K_S^0$ & 10.3 & 1.0 & 2.7 & $-18.5$ & 5.6 & 7.0 & 32.0 & 6.0 & 19.6 \\
$D^0\to K^*(1680)^+K^-$ & 17.1 & 2.3 & 2.4 & 151.6 & 8.3 & 7.3 & 26.1 & 6.0 & 24.0 \\
$D^0\to \rho (1700)^-\pi^+$ & 5.5 & 1.8 & 1.5 & 23.5 & 12.5 & 21.5 & 4.4 & 3.0 & 3.7 \\
$D^0\to K_0^*(1430)^+K^-$ & 1.5 & 0.3 & 0.7 & $-101.5$ & 18.9 & 8.7 & 3.1 & 1.2 & 3.1 \\
$D^0\to a_0(1450)^-\pi^+$ & 0.7 & 0.2 & 0.2 & -31.7 & 23.3 & 16.1 & 2.3 & 1.2 & 1.5 \\

\hline\hline\end{tabular}
\end{table*}

\begin{table*}[htbp]
\caption{The results of the Dalitz plot fit for the decay \dtosup. This is the best model including the neutral $K^*(892)^0$ resonance, suppressed model 1. The amplitudes, phases, and fit fractions are given. The first uncertainty is statistical, and the second systematic.\label{tab:ampsupres}}
\begin{tabular}[c]{lr@{$\pm$}c@{$\pm$}rr@{$\pm$}c@{$\pm$}rr@{$\pm$}c@{$\pm$}r}
\hline\hline
Component & \multicolumn{3}{c}{Amplitude} & \multicolumn{3}{c}{Phase ($^\circ$)} & \multicolumn{3}{c}{Fraction (\%)}\\
\hline

$D^0\to K^*(892)^-K^+$ & \multicolumn{3}{c}{1.0 (fixed)} & \multicolumn{3}{c}{0.0 (fixed)} & 20.4 & 2.1 & 0.8 \\
$D^0\to K^*(892)^0K_S^0$ & 0.4 & 0.1 & 0.0 & 109.9 & 10.8 & 3.7 & 3.9 & 1.5 & 0.4 \\
$D^0\to K_0^*(1430)^-K^+$ & 6.2 & 0.9 & 0.5 & $-114.4$ & 9.8 & 5.0 & 18.4 & 2.8 & 1.5 \\
$D^0\to a_0(1450)^+\pi^-$ & 3.1 & 0.5 & 0.4 & $-54.0$ & 8.6 & 6.4 & 15.8 & 4.1 & 2.1 \\
$D^0\to \rho (1700)^+\pi^-$ & 11.4 & 2.2 & 3.9 & 90.3 & 9.6 & 7.7 & 6.2 & 2.1 & 2.5 \\

\hline\hline\end{tabular}
\end{table*}

\begin{table*}[htbp]

\caption{The results of the Dalitz plot fit for the decay \dtosup. This is the model with the lowest $\chi^2$ per degree of freedom, suppressed model 2. The amplitudes, phases, and fit fractions are given. The first uncertainty is statistical, and the second systematic.\label{tab:ampsupres2}}
\begin{tabular}[c]{lr@{$\pm$}c@{$\pm$}rr@{$\pm$}c@{$\pm$}rr@{$\pm$}c@{$\pm$}r}
\hline\hline
Component & \multicolumn{3}{c}{Amplitude} & \multicolumn{3}{c}{Phase ($^\circ$)} & \multicolumn{3}{c}{Fraction (\%)}\\
\hline

$D^0\to K^*(892)^-K^+$ & \multicolumn{3}{c}{1.0 (fixed)} & \multicolumn{3}{c}{0.0 (fixed)} & 21.1 & 2.2 & 3.3 \\
$D^0\to K_0^*(1430)^-K^+$ & 8.0 & 0.9 & 0.7 & $-71.8$ & 13.4 & 9.3 & 31.5 & 6.2 & 8.0 \\
$D^0\to \rho (1700)^+\pi^-$ & 12.7 & 2.2 & 5.4 & 103.7 & 16.1 & 11.2 & 8.0 & 2.2 & 3.8 \\
$D^0\to a_0(1450)^+\pi^-$ & 1.8 & 0.4 & 0.3 & $-30.1$ & 14.7 & 10.9 & 5.8 & 2.3 & 1.6 \\
$D^0\to K_2^*(1430)^0K_S^0$ & 4.7 & 1.5 & 2.0 & $-104.1$ & 14.9 & 7.5 & 4.3 & 2.8 & 3.4 \\

\hline\hline\end{tabular}
\end{table*}

\section{Measurement of the ratio of branching fractions of suppressed and favored decays }
\label{sec:ratio}

The measurement to determine the ratio of branching fractions uses only \cleoiii\ data to avoid the complications of accounting for the quantum correlations present in the CLEO-c data. The ratio \braf is given by
\begin{equation}
\braf = \frac{N(D^0 \to K_S^0 K^+ \pi^-)}{N(D^0 \to K_S^0 K^- \pi^+)} \times \frac{\epsilon_{D^0 \to K_S^0 K^- \pi^+}}{\epsilon_{D^0 \to K_S^0 K^+ \pi^-}},
\end{equation}
where $N$ is the number of signal events for each decay, and $\epsilon$ is the efficiency.

The ratio of the efficiencies is not exactly unity as the distributions of the events over the Dalitz plot are different in the favored and suppressed decays and the efficiency over the Dalitz plot is not uniform. The ratio of the efficiencies is found  by taking the ratio of the integral over all of phase space for the default isobar models multiplied by the efficiency model, for the favored and suppressed decay. The efficiency ratio is determined to be 0.940 using favored model 1 and suppressed model 1. The uncertainty is assessed by altering the detector efficiency parametrization and using various alternative models arising from the fitting procedure that give a reasonable fit to data and leads to an uncertainty of $\pm$0.029 on the correction factor.
Using the tagged \cleoiii\ yields and the necessary corrections, \braf is measured to be 0.592 $\pm$ 0.044 $\pm$ 0.018, where the first uncertainty is statistical and the second systematic due to uncertainties in the correction factors. This measurement is consistent, and more precise than $0.79\pm 0.19$, the average of previous measurements~\cite{PDG}. The measurement for \braf is also required in the restricted kinematic region of 100 \mevcc around the $K^{*}(892)^\pm$ resonance for the coherence factor measurement and is here determined to be $\mathcal{B}_{K^*K} = 0.356\pm0.034\pm0.007$.

\section{Coherence factor and average strong--phase difference measurement}
\label{sec:coh}

\subsection{Quantum--Correlated $D$ Decays}
The coherence factor \cho\ and the average strong--phase difference \spd\ can be determined using double--tagged $D^0\bar{D^0}$ pairs from the $\psi(3770)$ resonance produced in $e^+ e^-$ collisions at $\sqrt{s}=3770$ MeV. The two mesons are produced in a $C$--odd eigenstate and therefore the decays of the two $D$ mesons are quantum--correlated. There are three categories of double tags used. The first of these are the flavor tags, where the samples of double tags are split into two classes: those where the tag kaon and the kaon from the signal side have the same sign of charge (SS) and those where the two kaons have opposite sign of charge (OS). The second category of double tags are those where the tag $D$ meson decays to a $CP$ eigenstate, which can either be even or odd. The final category of double tags used are those where the tag $D$ meson decays to $K^0_{S,L}\pi^+\pi^-$. Each category of double tags has a different sensitivity to \cho\ and \spd.
 
The rate for the two neutral $D$ mesons produced from $\psi(3770)$ where one decays to the \kskpi\ final state and the other decays to any final state $G$, $\Gamma(\kskpi|G)$,  is given by~\cite{coh}
\begin{widetext}
\begin{eqnarray}
\Gamma(\kskpi|G) &=& \Gamma_0 \int \int \int |\mathcal{A}_{K_S^0K\pi}(m^2_{K_S^0K},m^2_{K\pi}) \mathcal{A}_{\overline{G}}(z) - \mathcal{A}_{\overline{K_S^0K\pi}}(m^2_{K_S^0K},m^2_{K\pi}) \mathcal{A}_{{G}}(z)|^2 dm^2_{K_S^0K}dm^2_{K\pi}dz \nonumber \\
&=& \Gamma_0 [A_{K_S^0K\pi}^2 A_{\overline{G}}^2 + A_{\overline{K_S^0K\pi}}^2A_{G}^2 - 2R_{K_S^0K\pi}R_GA_{K_S^0K\pi}A_{\overline{K_S^0K\pi}}A_GA_{\overline{G}} \cos(\delta^G_D-\delta^{K_S^0K\pi}_D)]\label{eq:4},
\end{eqnarray}
\end{widetext}
where $z$ describes the multi--body phase space for final state $G$, $R_G$ and $\delta_G$ are the coherence factor and average strong--phase difference for the final state $G$ as defined in an analogous way to that given in~\equationref{eq:defcho}, and $\Gamma_0$ is the normalization factor. Specifying that the tag $D$ meson decays to either a flavor tag or a $CP$ tag, ~\equationref{eq:4} simplifies to the following relations:
\begin{widetext}
\begin{eqnarray}
\Gamma({K_S^0K\pi|CP^{\pm}}) &=& \Gamma_0A_{K_S^0K\pi}^2A_{CP}^2[1 + (r_D^{K_S^0K\pi})^2 - 2\lambda_{\pm}r_D^{K_S^0K\pi}\cho \cos\spd],\label{eq:ratecp}\\
\Gamma(K_S^0K\pi|F^{SS}) &=& \Gamma_0 A^2_{K_S^0K\pi} A^2_{F} [ (r^{K_S^0K\pi}_D)^2+(r^{F}_D)^2 - 2r_D^{F}r^{K_S^0K\pi}_D R_{K_S^0K\pi}R_F\cos(\delta_D^{F}-\delta_D^{K_S^0K\pi})],\\
\Gamma(K_S^0K\pi|F^{OS}) &=& \Gamma_0 A^2_{K_S^0K\pi} A^2_{F} [1+ (r_D^{K_S^0K\pi} r^{F}_D)^2 - 2r_D^{F}r^{K_S^0K\pi}_D R_{K_S^0K\pi}R_F\cos(\delta_D^{F}-\delta_D^{K_S^0K\pi})],\label{eq:rateOS}
\end{eqnarray}
\end{widetext}
where $F$, the flavor tag is either $K^\mp\pi^\pm$, $K^\mp\pi^\pm\pi^+\pi^-$, or $K^\mp\pi^\pm\pi^0$, the superscript $SS$ or $OS$ denotes the charge relation between the signal and tag kaon, and $CP$ denotes a $CP$ eigenstate with eigenvalue $\lambda_{\pm} = \pm 1$. The quantity $r_D^F$ is analogous to~\equationref{eq:defrd} using the DCS and CF amplitudes for the given flavor tag. These rate expressions ignore $CP$ violation in $D$ decays, which is well motivated theoretically and experimentally known to be very small~\cite{noCPV, lhcbdeltaAg}. 
These rate equations allow for the number of expected events in a given data sample to be calculated for each category of double tag. The left hand side can be replaced with the efficiency--corrected background--subtracted signal double--tag yield, $S(K_S^0K\pi|G)$, where
\begin{equation}
 S(K_S^0K\pi|G)= N(K_S^0K\pi|G)/\varepsilon(K_S^0K\pi|G),
\end{equation}
where $  N(K_S^0K\pi| G)$ is the background--subtracted yield of double tags where one $D$ meson decays to $K_S^0K\pi$ and the other to final state $G$, and $\varepsilon(K_S^0K\pi|G)$ is the efficiency to select that particular double tag. Ratios of the rate equations are taken to remove the dependence on the normalization factor and $A_{K_S^0K\pi}$. This leads to the useful observables $\nu^F$ and $\kappa^\pm$ defined as
\begin{widetext}
\begin{eqnarray}
\nu^{F} &\equiv& \frac{S({K_S^0K\pi}|F^{SS})}{S({K_S^0K\pi}|F^{OS})} = \frac{(r_D^{F})^2 + (r_D^{K_S^0K\pi})^2 -2r_D^{F}r_D^{K_S^0K\pi}R_F[p\cos(\delta^{F}) +q\sin(\delta^{F})]}{1+(r_D^{F})^2(r_D^{K_S^0K\pi})^2 -2r_D^{F}r_D^{K_S^0K\pi}R_F[p\cos(\delta^{F}) +q\sin(\delta^{F})]}\label{eq:nu},
\end{eqnarray}
\begin{eqnarray}
\begin{split}
\kappa^\pm &\equiv \frac{S({K_S^0K\pi}|CP^\pm) \mathcal{B}(D \to K\pi)}{S({K_S^0K\pi}|K\pi^{SS}) \mathcal{B}(D \to CP)}\\
&= \frac{1 + (r_D^{K_S^0K\pi})^2 - 2\lambda_\pm r_D^{K_S^0K\pi}p }{(1-\lambda_{\pm}y){(r_D^{K\pi})^2 + (r_D^{K_S^0K\pi})^2 - 2r_D^{K\pi}r_D^{K_S^0K\pi} [p\cos(\delta^{K\pi}) +q\sin( \delta^{K\pi})]}},
\label{eq:kappa}
\end{split}
\end{eqnarray}
\end{widetext}
where
\begin{eqnarray}
p&\equiv& \cho\cos\spd, \\
q&\equiv& \cho \sin\spd,
\end{eqnarray}
and $y$ is the charm mixing parameter given by $y=(\Gamma_2-\Gamma_1)/2\Gamma$ where $\Gamma_2$ and $\Gamma_1$, are the lifetimes of the two $D$ meson mass eigenstates and $\Gamma$ is the mean of $\Gamma_2$ and $\Gamma_1$. The branching fractions are included using the following relations: $\mathcal{B}(D^0 \to CP^\pm) = A_{CP}^2(1 \mp y)$ and $\mathcal{B}(D^0 \to K^-\pi^+)=A_{K^-\pi^+}^2$. In the case of the $\kappa$ observable the yields and branching fractions are combined so that the true value of the $\kappa^+$ ($\kappa^{-}$) observable is the same for all $CP$ even (odd) tags.

The other tag group which also provides sensitivity to \cho\ and \spd\  is $K^0_{S,L}\pi^+\pi^-$. The Dalitz plot of $K^0_{S,L} \pi^+\pi^-$ can be described by two coordinates $m^2_+$ and $m^2_{-}$, where $m_+$($m_-$) is the invariant mass of the $K^0_{S,L}\pi^+$($K^0_{S,L}\pi^-)$ pair. The Dalitz plot can be divided into bins where the $i^{\mathrm{th}}$ bin has  $m^2_+>m^2_-$, and the symmetric bin in the region where $m^2_+<m^2_-$ is labeled the $-i^{\mathrm{th}}$ bin. For each  point the phase difference, $\Delta \delta_D$, is defined as $\Delta \delta_D \equiv \delta_D(m_+^2, m_-^2)- \delta_D(m_-^2, m_+^2)$, where  $\delta_D(m_+^2, m_-^2)$ is the phase of the $D^0$ decay at that point. In each bin the amplitude--weighted averages of $\cos(\Delta(\delta_D))$ and $\sin(\Delta(\delta_D))$ over each Dalitz plot bin, denoted $c_i$ and $s_i$ respectively can be determined. These $c_i$ and $s_i$ measurements have been performed~\cite{kspipipap}, together with the equivalent parameters for $K^0_L\pi^+\pi^-$ decays, $c_i'$ and $s_i'$. In addition the flavor tag yields in each bin, $L_i$ and $L_{-i}$ ($L_i'$ and $L_{-i}'$) for $K_S^0\pi^+\pi^-$ ($K_L^0\pi^+\pi^-$) have also been determined. The binning choice used in the current analysis is that of the equal regions in $\Delta(\delta_D)$ as defined in ~\cite{kspipipap}.

In quantum--correlated decays of $D$ mesons at $\sqrt{s}=3770$ \mev, the yield of the double tag where the tag $D$ meson decays to $K_{S,L}\pi^+\pi^-$ in the $i^{\mathrm{th}}$ bin is given by
\begin{widetext}
\begin{eqnarray}
 S(K_S^0\pi^+\pi^-|K_S^0K^+\pi^-)_i &=& H_S(L_i + (r_D^{K_S^0K\pi})^2 L_{-i} - 2r_D^{K_S^0K\pi}\sqrt{(L_i L_{-i})}[c_i p - s_i q]),\label{eq:Kspipi}\\
S(K_L^0\pi^+\pi^-|K_S^0K^+\pi^-)_i &=&  H_L(L_i' + (r_D^{K_S^0K\pi})^2 L_{-i}' + 2r_D^{K_S^0K\pi}\sqrt{(L_i' L_{-i}')}[c_i'p - s_i'q])\label{eq:Klpipi}, 
\end{eqnarray}
\end{widetext}
where $H_S$ and $H_L$ are normalization factors, and $S$ is the efficiency--corrected signal yield in the $i^{\mathrm{th}}$ bin of the Dalitz plot. The inclusion of the $K_{S,L}^0\pi\pi$ tags is necessary in the measurement as the flavor and $CP$ tags on their own provide limited sensitivity to $q$. 

\subsection{Measurement of observables}

The background--subtracted signal yields of each double tag are given in ~\secref{sec:datacleoc}. To determine the efficiency--corrected yields it is necessary to know the efficiency for each double tag. The double--tag reconstruction efficiencies are determined from separate simulated signal samples for each final state. The resonant decay structure is not simulated for any of the $D^0$ meson decays because no significant differences are observed between the overall efficiency determined with flat simulation and the resonant decay simulation. In each final state 20,000 events are generated, except for the $K^0_{S,L}\pi^+\pi^-$ tags where 200,000 events are generated so that the statistical uncertainty on the determined efficiency remains small in each bin of the Dalitz plot. The determined efficiencies are corrected for known small differences between data and simulation due to $\pi^0$ reconstruction and particle identification efficiencies. The double tag reconstruction efficiencies range from $20\%$ for $\pi^+\pi^-$ to $2\%$ for the $K_S^0\eta'$ tag. The variation in efficiency from bin to bin in either $K_S^0\pi^+\pi^-$ or $K_L^0\pi^+\pi^-$ tags is less than $2\%$. The effect of the resonant structure on the determined efficiency is taken into account as a systematic uncertainty. 

As the resonant decays of either $D$ meson are not simulated the measured efficiencies for the SS flavor tags and OS flavor double tags are identical. The systematic uncertainty arising from any difference that does exist is considered separately. Hence the $\nu$ values defined in ~\equationref{eq:nu} are simply the ratios of the background--subtracted signal yields given in ~\tabref{tab:pseudoyields} and are  0.66 $\pm$ 0.10, 0.63 $\pm$ 0.08, and 0.54 $\pm$ 0.06 for $\nu^{K\pi}$, $\nu^{K\pi\pi\pi}$, and $\nu^{K\pi\pi^0}$, respectively. The values of the $S(K_S^0K\pi|CP)$ and $S(K_S^0K\pi|K\pi^{SS})$ yields are listed in Tables~\ref{tab:syields1} and~\ref{tab:syields2}, respectively. The values of the $S(K_S^0K\pi|K^0\pi\pi)$ yields are listed in Tables~\ref{tab:syields3} and~\ref{tab:syields4}.

\begin{table*}[thpb]
\begin{center}

\caption{Efficiency--corrected yields for double tags with no $K_L^0$ in final state. The uncertainties take into account the statistical uncertainties on the uncorrected yields only. \label{tab:syields1}}
\begin{tabular}{lccccccccc}
\hline \hline
Tag & $KK$  & $\pi\pi$  & $K_S^0\pi^0\pi^0$  & $K_S^0\pi^0$  & $K_S^0\omega$  & $K_S^0\eta ( \gamma\gamma )$  & $K_S^0\eta(\pi\pi\pi^0)$  & $K_S^0\eta '$  & $K\pi^{SS}$ \\
$S(K_S^0K\pi|G)$ & 10.7$\pm$16.0 & 16.1$\pm$15.1 & 274.5$\pm$89.8 & 305.0$\pm$65.7 & 293.7$\pm$89.9 & 93.8$\pm$36.1 & 7.5$\pm$18.6 & 52.1$\pm$62.6 & 989.5$\pm$110.5\\
\hline \hline
\end{tabular}

\end{center}
\end{table*}

\begin{table*}[thpb]
\begin{center}
\caption{Efficiency--corrected signal yields for the $CP$ tags with a $K_L^0$. The uncertainties take into account the statistical uncertainties on the uncorrected yields only. \label{tab:syields2}}
\begin{tabular}{lccccc}
\hline \hline
 Tag& $K_L^0\pi^0$  & $K_L^0\omega$  & $K_L^0\eta$  & $K_L^0\eta '$  & $K_L^0\pi^0\pi^0$ \\
$S(K_S^0K\pi|G)$ & 171.6$\pm$45.0 & 58.0$\pm$53.0 & $-1.0$$\pm$13.1 & 9.1$\pm$34.0 & 363.2$\pm$262.3\\
\hline \hline
\end{tabular}
\end{center}
\end{table*}

\begin{table*}[thpb]
\begin{center}
\caption{Efficiency--corrected signal yields in the $K_S^0\pi^+\pi^-$ bins. The uncertainties take into account the statistical uncertainties on the uncorrected yields only.\label{tab:syields3}}
\begin{tabular}{lcccccccc}
\hline \hline
Bin$_i$ & 1 & 2 & 3 & 4 & 5 & 6 & 7 & 8\\
$S(K_S^0K\pi|K_S^0\pi^+\pi^-)_i$ & 55.9$\pm$26.0 & 17.3$\pm$15.0 & 18.0$\pm$14.4 & 30.8$\pm$18.1 & 70.8$\pm$28.9 & 8.8$\pm$10.0 & 20.1$\pm$16.8 & 37.2$\pm$20.2\\
\\
Bin$_i$ & $-1$ & $-2$ & $-3$ & $-4$ & $-5$ & $-6$ & $-7$ & $-8$\\
$S(K_S^0K\pi|K_S^0\pi^+\pi^-)_i$ & 42.0$\pm$24.1 & 40.5$\pm$22.2 & 53.6$\pm$24.9 & 35.3$\pm$19.4 & 81.4$\pm$29.6 & 43.2$\pm$21.9 & 5.3$\pm$11.1 & 76.9$\pm$29.6\\
\hline \hline
\end{tabular}
\end{center}
\end{table*}

\begin{table*}[thpb]
\begin{center}
\caption{Efficiency--corrected signal yields in the $K_L^0\pi^+\pi^-$ bins. The uncertainties take into account the statistical uncertainties on the uncorrected yields only. \label{tab:syields4}}
\begin{tabular}{lcccccccc}
\hline \hline
Bin$_i$ & 1 & 2 & 3 & 4 & 5 & 6 & 7 & 8\\
$S(K_S^0K\pi|K_L^0\pi^+\pi^-)_i$ & 123.6$\pm$28.4 & 28.7$\pm$14.5 & 28.5$\pm$13.8 & 4.0$\pm$6.8 & 20.7$\pm$13.2 & 18.2$\pm$12.2 & 107.0$\pm$25.2 & 106.5$\pm$25.5\\
\\
Bin$_i$ & $-1$ & $-2$ & $-3$ & $-4$ & $-5$ & $-6$ & $-7$ & $-8$\\
$S(K_S^0K\pi|K_L^0\pi^+\pi^-)_i$ & 128.5$\pm$29.4 & 48.5$\pm$17.5 & 20.1$\pm$11.5 & 9.5$\pm$9.2 & 14.3$\pm$12.3 & 21.3$\pm$13.0 & 53.5$\pm$18.7 & 58.0$\pm$19.1\\
\hline \hline
\end{tabular}
\end{center}
\end{table*}

To determine $\kappa$ for each $CP$ tag the branching fractions are taken from~\cite{PDG}. All subsidiary branching fractions such as $K_S^0\to \pi^+\pi^-$ are included in the calculation. The uncertainties on the $\kappa$ value include a statistical component from the double--tag yield, and systematic uncertainties arising from the uncertainties in the difference between Monte Carlo and data, the scale factors used to determine background yields, the uncertainties of branching fractions used, and the yield of the $K\pi^{SS}$ double tag. A number of these uncertainties are correlated among some or all of the observables. 

The systematic uncertainties due to branching fractions, the background scale factors, and  $N(K_S^0K\pi|K\pi^{SS})$ are determined by shifting the default value of the branching fraction, scale factor or $N(K_S^0K\pi|K\pi^{SS})$ by one standard deviation and taking the shift on the observable as the systematic uncertainty.  There are systematic uncertainties due to uncorrected differences between simulation and data tracking and reconstruction efficiencies. Relative uncertainties of 2$\%$ for each tag containing a $\pi^0$,  $0.6\%$ per kaon track, $0.3\%$ per charged pion track, and $0.8\%$ for each $CP$ tag containing a $K_S^0$ are applied, determined from studies of simulated and real data. The shift in $\kappa$ due to the altered efficiencies is taken as the systematic uncertainty. The systematic effect due to non--uniform efficiency over the Dalitz plot is treated separately and discussed in ~\secref{sec:moresyst}. 

To combine the systematic uncertainties an error matrix is constructed where the diagonal elements are separate uncertainties combined in quadrature, and the off--diagonal elements are the sum of all $\sigma_j\sigma_k$ where the effect of the uncertainty is correlated between observables $j$ and $k$. The systematic uncertainties on the $\kappa$ observables are given in Tables~\ref{tab:sysCPeven} and ~\ref{tab:sysCPOdd}, and the central values, statistical, and total systematic uncertainties summarized in Tables~~\ref{tab:valsysCPeven} and ~\ref{tab:valsysCPOdd}. The systematic uncertainties are smaller than the statistical uncertainties. The leading systematic uncertainties arise due to uncertainties on the branching ratios and the estimated background yields. The average values for $\kappa^+$ and $\kappa^-$, taking into account systematic uncertainties and correlations, are 0.56 $\pm$ 0.22 and 2.64 $\pm$ 0.67, respectively. These averages are determined by minimizing the $\chi^2$ between the measured and average $\kappa$ values. The $\chi^2$/dof is 12.0/11 which indicates a good overall compatibility of the individual measurements. The significant difference between $\kappa^+$ and $\kappa^-$ values indicates the presence of quantum correlations and a non--zero value for the quantity $p$.  In comparison with the incoherent expectation there is an excess of $CP$--odd tags and a reduction in the yield of $CP$--even tags, and hence even the low yields of the $CP$--even tags carry significant information. The error matrix for the $\kappa$ observables is given in Appendix B.  

 The systematic uncertainty on the $\nu$ observables is very small as there are no contributions from efficiencies or branching fractions. The only systematic uncertainty considered is that on background yields which partly cancels in the ratio.  In the case of the $K^0_{S,L}\pi^+\pi^-$ tags the observable is simply the efficiency--corrected background--subtracted yield. The background yield uncertainty is calculated in the same way as for the $CP$ tags and the correlations between the uncertainties for the $K^0\pi^+\pi^-$ tags and the $CP$ tags are determined, and included in the error matrix. Global changes to the efficiencies, those which are the same for all bins, are not relevant as any effect on the efficiency--corrected yield will simply be absorbed into the normalization factors in the fit. The yields are determined per bin in the Dalitz plot, which gives rise to the possibility that detector resolution will lead to bin migration. This is not considered here as bin migration effects are already taken into account in the uncertainties on the measured values of $c_i$ and $s_i$.

\begin{table*}[thpb]
\begin{center}
\caption{The systematic uncertainties on the $\kappa^+$ values.\label{tab:sysCPeven}}
\begin{tabular}{lccccccc}
\hline \hline

 & $KK$  & $\pi\pi$  & $K_L^0\pi^0$  & $K_L^0\omega$  & $K_L^0\eta$  & $K_L^0\eta '$  & $K_S^0\pi^0\pi^0$ \\
\hline
$CP$ tag branching fraction & 0.004 & 0.017 & 0.095 & 0.077 & 0.005 & 0.060 & 0.553\\
 $D^0 	\to K\pi$ branching fraction & 0.003 & 0.012 & 0.019 & 0.007 & 0.001 & 0.006 & 0.042\\
Background scaling ratio & 0.057 & 0.196 & 0.062 & 0.028 & 0.128 & 0.090 & 0.015\\
$K\pi$ tag yield& 0.028 & 0.114 & 0.183 & 0.064 & 0.006 & 0.058 & 0.408\\
$CP$ tag tracking and reconstruction  & 0.002 & 0.004 & 0.029 & 0.010 & 0.000 & 0.002 & 0.070\\
$K\pi$ tracking and reconstruction & 0.001 & 0.006 & 0.010 & 0.003 & 0.000 & 0.003 & 0.022\\
\\
Total syst. & 0.064 & 0.228 & 0.218 & 0.105 & 0.129 & 0.122 & 0.693\\
\hline \hline

\end{tabular}
\end{center}
\end{table*}

\begin{table*}[thpb]
\begin{center}
\caption{The central values $\kappa^+$ and the statistical and systematic uncertainties.\label{tab:valsysCPeven}}
\begin{tabular}{lccccccc}
\hline \hline

 & $KK$  & $\pi\pi$  & $K_L^0\pi^0$  & $K_L^0\omega$  & $K_L^0\eta$  & $K_L^0\eta '$  & $K_S^0\pi^0\pi^0$ \\
\hline

Central $\kappa$ value & 0.219 & 0.909 & 1.456 & 0.512 & $-0.049$ & 0.459 & 3.247\\ 
\\
Statistical uncertainty & 0.328 & 0.855 & 0.382 & 0.467 & 0.665 & 1.702 & 1.063\\
Total systematic uncertainty & 0.064 & 0.228 & 0.218 & 0.105 & 0.129 & 0.122 & 0.693\\
\\
Total stat. + syst. & 0.334 & 0.885 & 0.440 & 0.479 & 0.677 & 1.706 & 1.269\\
\hline \hline

\end{tabular}
\end{center}
\end{table*}

\begin{table*}[thpb]
\begin{center}
\caption{The systematic uncertainties on the $\kappa^-$ values.\label{tab:sysCPOdd}}
\begin{tabular}{lcccccc}
\hline \hline
 & $K_S^0\pi^0$  & $K_S^0\omega$  & $K_S^0\eta ( \gamma\gamma )$  & $K_S^0\eta(\pi\pi\pi^0)$  & $K_S^0\eta '$  & $K_L^0\pi^0\pi^0$ \\
\hline
Tag branching fraction & 0.145 & 0.571 & 0.773 & 0.117 & 0.498 & 0.499\\
 $D^0 	\to K\pi$ branching fraction & 0.040 & 0.049 & 0.089 & 0.013 & 0.049 & 0.038\\
Background scaling ratio & 0.004 & 0.015 & 0.010 & 0.030 & 0.008 & 0.075\\
$K\pi$ tag yield& 0.390 & 0.476 & 0.873 & 0.131 & 0.483 & 0.368\\
$CP$ tag tracking and reconstruction & 0.067 & 0.083 & 0.063 & 0.023 & 0.038 & 0.057\\
$K\pi$ tracking and reconstruction& 0.021 & 0.025 & 0.047 & 0.007 & 0.026 & 0.020\\
\\
Total syst. & 0.424 & 0.750 & 1.172 & 0.180 & 0.697 & 0.628\\
\hline
\hline

\end{tabular}
\end{center}
\end{table*}

\begin{table*}[thpb]
\begin{center}
\caption{The central values $\kappa^-$ and the statistical and systematic uncertainties.\label{tab:valsysCPOdd}}
\begin{tabular}{lcccccc}
\hline \hline
 & $K_S^0\pi^0$  & $K_S^0\omega$  & $K_S^0\eta ( \gamma\gamma )$  & $K_S^0\eta(\pi\pi\pi^0)$  & $K_S^0\eta '$  & $K_L^0\pi^0\pi^0$ \\
\hline
Central $\kappa$ value & 3.102 & 3.791 & 6.949 & 1.043 & 3.841 & 2.928\\
\\
Statistical uncertainty & 0.669 & 1.160 & 2.676 & 2.595 & 4.614 & 2.114\\
Total systematic uncertainty& 0.424 & 0.750 & 1.172 & 0.180 & 0.697 & 0.628\\
\\
Total stat. + syst. & 0.792 & 1.382 & 2.921 & 2.602 & 4.667 & 2.205\\
\hline
\hline

\end{tabular}
\end{center}
\end{table*}

\subsection{Combination and further systematic uncertainties}
\label{sec:moresyst}
To determine the coherence factor and average strong--phase difference the information on the observables is combined in a $\chi^2$ fit to determine the best--fit values for $p$ and $q$. To take into account all the uncertainties and their correlations the $\chi^2$ is given by
\begin{equation}
\chi^2= \Delta\rho{\bf E^{-1}}\Delta\rho^T,
\end{equation}
where $\Delta \rho$ is a vector of differences between the expected and observed values of the observables and ${\bf{E}}$ is the error matrix derived from the total uncertainties and their correlations. The expected values are computed using either ~\equationref{eq:nu},~\equationref{eq:kappa},~\equationref{eq:Kspipi}, or ~\equationref{eq:Klpipi} as appropriate for a given value of $p$ and $q$. To calculate the expected values a number of external inputs are required. These external inputs are free parameters in the fit but are Gaussian--constrained to their measured values, which takes into account any systematic uncertainties arising from the measured central values of the external inputs. Correlations between the uncertainties of the external parameters are taken into account where known. The values of $\delta^{K\pi}$, $r_D^{K\pi}$, $x$, and $y$ and their uncertainties and correlations are taken from~\cite{HFAG}. The information on  $c_i^{(')}$, $s_i^{(')}$ is from~\cite{kspipipap}, and the coherence information on other final states from~\cite{NewCoh}. The value of $r_D^{K_S^0K\pi}$ used is determined using the measured value of \braf in ~\cite{LHCbOlli}\footnote{The originally reported results in v2 of this preprint used the branching fractions measured within. On correction of equations (25), (26), (28) in 2016 the more precise value measured by the LHCb collaboration above is used.} and the following relation between \braf and the $D$ mixing parameters~\cite{dmix}:
\begin{equation}
\braf = \frac{(r_D^{K_S^0K\pi})^2 + (yp-xq)r_D^{K_S^0K\pi}}{1+ (yp+xq)r_D^{K_S^0K\pi}},
\end{equation}  
where the quadratic and higher order terms in $y$ and $x$ are negligible.

 The $\chi^2$ is minimized to determine the best--fit values of $p$ and $q$. The normalization factors $H_S$ and $H_L$ of Eqs.~(\ref{eq:Kspipi}) and (\ref{eq:Klpipi}) are also free parameters in the fit. The best--fit values are $p$ = 0.701 $\pm$ 0.078 and $q$ = 0.001 $\pm$ 0.192. The $\chi^2$/dof of the fit is 47.0/41. The performance of the fit is tested using simple simulation. The simple simulation generates observable values based on input values of \cho\ and \spd, where the observable value for each tag is smeared by the experimentally observed uncertainties. The studies using the simple simulation confirm that the use of tag yields that are consistent with zero do not cause bias and that tags with low yields do add sensitivity to the overall measurement.

The efficiencies are determined from Monte Carlo where the decay products are generated uniformly over phase space. However the efficiency is not uniform over the Dalitz plot, and furthermore as the distribution of events over the Dalitz plot is different depending on the tag due to quantum correlations a further systematic uncertainty must be considered. This uncertainty is determined by using the isobar models described in ~\secref{sec:ampdefs}. The shifts in the values of $p$ and $q$ are computed using ~\equationref{eq:defcho} and inserting either a flat efficiency or the CLEO-c efficiency model for a variety of isobar models including the reported ones. The largest shifts observed between the flat and CLEO-c efficiency model are 0.016 for $p$ and 0.002 for $q$ which are small in comparison to the measured uncertainty.  Hence these shifts are assigned as the systematic uncertainty from non--uniform efficiency. 

\begin{figure}[htbp]

\begin{center}
\includegraphics[width=0.99\columnwidth]{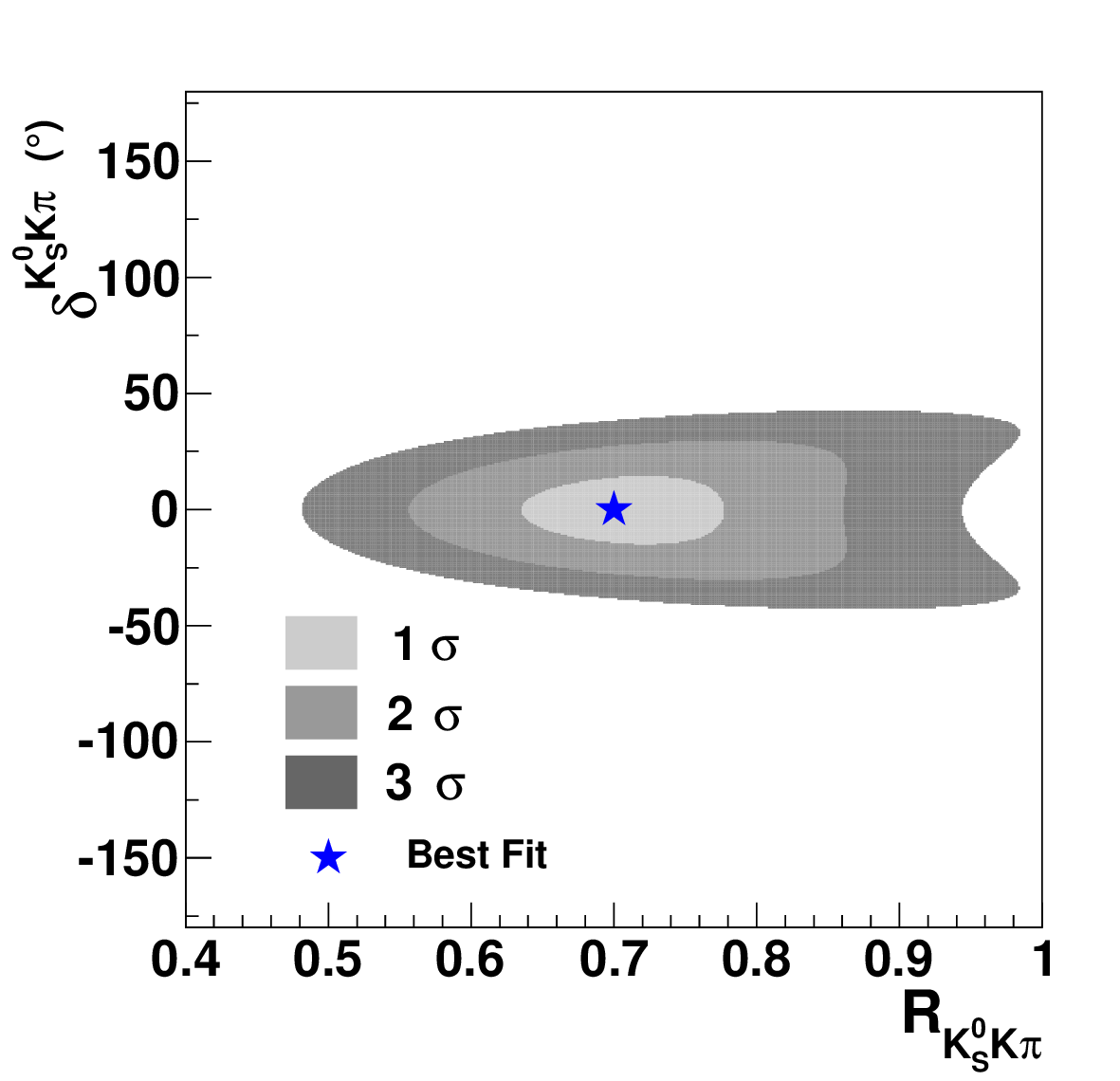}
\caption{The best--fit point for \cho\ and \spd\ measured over the whole Dalitz plot, and the regions enclosing 1, 2, and 3 standard deviations from that point. \label{fig:contfull}}
\end{center}
\end{figure}

 As $p$ is positive and \cho\ is defined to be between zero and one, the solution for \spd\ lies in the range from $-90^\circ$ to 90$^\circ$. This leads to \cho\ = $0.70\pm 0.08$ and \spd\ = (0.1$\pm 15.7)^\circ$. Figure~\ref{fig:contfull} shows regions of (\cho, \spd) parameter space from the fit consistent with one, two, and three standard deviations from the best--fit point. The likelihood is computed as $\mathcal{L}=e^{-(\chi^2-\chi^2_{\mathrm{min}})/2}$ at a point in parameter space; the fit is repeated at each point with the values of $p$ and $q$ fixed. The isobar models also give predictions of the coherence factor via ~\equationref{eq:defcho}. The isobar models predict a coherence factor of 0.72 (0.79) when the central values of favored and suppressed models 1 (2) are used. The predictions are consistent with the measured value. 

\subsection{Coherence factor in a restricted region}
\label{sec:cohres}
 Around dominant resonances occurring in both Dalitz plots,  the coherence factor is expected to be close to one, as this situation approximates the two--body decay which is fully coherent. Higher coherence in principle provides increased sensitivity to $\gamma$. Therefore the analysis is repeated using only the data where the invariant mass of the $K_S^0\pi$ combination lies within 100 \mevcc\ of the $K^{*}(892)^\pm$ mass. The analysis could, in principle, also be performed in the region outside this invariant mass window; however the data are insufficient for this complementary study. The efficiencies for each double tag are recalculated for the restricted region. The efficiency--corrected yields for the $CP$ tags and $K\pi^{SS}$ tag in this restricted region are given in Tables~\ref{tab:kscpyields1} and~\ref{tab:kscpyields2}. The efficiency--corrected yields for $K^0_{S,L}\pi^+\pi^-$ tags are given in ~\tabref{tab:kscpyields3} and ~\tabref{tab:kscpyields4}. The expected yield of the $K_L^0\eta'$ tag is very small and is not used. The error matrix relating the uncertainties for the $\kappa$ observables is given in Appendix B. In the restricted region, the values of the $\nu$ variables are 0.40$\pm$0.08, 0.35$\pm$0.06 and 0.35$\pm$0.05 for $\nu^{K\pi}$, $\nu^{K\pi\pi\pi}$, and $\nu^{K\pi\pi^0}$, respectively.

\begin{table*}[thpb]
\begin{center}
\caption{Efficiency--corrected yields in the restricted region for double tags with no $K_L^0$ in final state.\label{tab:kscpyields1}}
\begin{tabular}{lccccccccc}
\hline \hline
Tag & $KK$  & $\pi\pi$  & $K_S^0\pi^0\pi^0$  & $K_S^0\pi^0$  & $K_S^0\omega$  & $K_S^0\eta ( \gamma\gamma )$  & $K_S^0\eta(\pi\pi\pi^0)$  & $K_S^0\eta '$  & $K\pi$ SS \\
\hline
$S(K_S^0K\pi|G)$& 6.0$\pm$14.2 & 2.0$\pm$11.1 & 43.7$\pm$37.8 & 233.1$\pm$57.9 & 199.9$\pm$68.7 & 47.8$\pm$28.2 & 10.1$\pm$18.1 & 83.0$\pm$87.5 & 288.9$\pm$48.7\\
\hline \hline
\end{tabular}
\end{center}
\end{table*}

\begin{table}[thpb]
\begin{center}
\caption{Efficiency--corrected signal yields in the restricted region for the $CP$ tags with a $K_L^0$. \label{tab:kscpyields2}}
\begin{tabular}{lcccc}
\hline \hline
Tag & $K_L^0\pi^0$  & $K_L^0\omega$  & $K_L^0\eta$  & $K_L^0\pi^0\pi^0$ \\
\hline
$S(K_S^0K\pi|G)$ & 116.2$\pm$37.1 & 25.9$\pm$38.9 & $-4.7$$\pm$8.5 & 186.4$\pm$194.0\\
\hline \hline
\end{tabular}
\end{center}
\end{table}

\begin{table*}[thpb]
\begin{center}
\caption{Efficiency--corrected signal yields in the restricted region in the $K_S^0\pi^+\pi^-$ bins.\label{tab:kscpyields3}}
\begin{tabular}{lcccccccc}
\hline \hline
Bin & 1 & 2 & 3 & 4 & 5 & 6 & 7 & 8\\
$S(K_S^0K\pi|K_S^0\pi^+\pi^-)_i$ & 40.3$\pm$36.8 & 48.3$\pm$35.8 & 19.5$\pm$23.6 & 49.3$\pm$35.8 & 128.2$\pm$60.4 & 25.1$\pm$27.5 & 54.6$\pm$44.3 & 18.8$\pm$23.8\\
\\
Bin & $-1$ & $-2$ & $-3$ & $-4$ & $-5$ & $-6$ & $-7$ & $-8$\\
$S(K_S^0K\pi|K_S^0\pi^+\pi^-)_i$ & 65.7$\pm$43.7 & 45.3$\pm$38.4 & 59.0$\pm$40.0 & 23.2$\pm$25.4 & 108.6$\pm$50.3 & 54.8$\pm$39.8 & 16.3$\pm$21.8 & 115.7$\pm$59.2\\
\hline \hline
\end{tabular}
\end{center}
\end{table*}

\begin{table*}[thpb]
\begin{center}
\caption{Efficiency--corrected signal yields in the restricted region in the $K_L^0\pi^+\pi^-$ bins.\label{tab:kscpyields4}}
\begin{tabular}{lcccccccc}
\hline \hline
Bin & 1 & 2 & 3 & 4 & 5 & 6 & 7 & 8\\
$S(K_S^0K\pi|K_L^0\pi^+\pi^-)_i $& 201.0$\pm$54.9 & 23.3$\pm$22.2 & 60.7$\pm$29.2 & $-9.4$$\pm$14.0 & 25.6$\pm$21.3 & $-1.2$$\pm$13.0 & 156.2$\pm$45.8 & 176.2$\pm$48.5\\
\\
Bin & $-1$ & $-2$ & $-3$ & $-4$ & $-5$ & $-6$ & $-7$ & $-8$\\
$S(K_S^0K\pi|K_L^0\pi^+\pi^-)_i$ & 242.5$\pm$59.6 & 79.2$\pm$33.0 & 26.5$\pm$19.4 & 2.4$\pm$12.6 & 29.0$\pm$22.5 & 12.9$\pm$19.8 & 86.4$\pm$34.7 & 65.2$\pm$29.6\\
\hline \hline
\end{tabular}
\end{center}
\end{table*}

 The value of $\mathcal{B}_{K^*K}$ that is used in the analysis is the one calculated for the restricted region in~\cite{LHCbOlli}. In the restricted kinematic region case, a further systematic uncertainty arises due to detector resolution as the candidate events may migrate across the defined boundary in the Dalitz plot. To account for this migration simulated signal data samples are generated using the resonant models determined in the amplitude analysis. For the $CP$ tags and the $K^0_{S,L}\pi^+\pi^-$ tags the quantum correlation between the tag and signal side is emulated. Candidate events can migrate into the restricted region, or they can migrate out. There is no systematic uncertainty if the migration in both directions is equal, but the presence of the resonance means that a larger fraction of events will migrate out of the restricted region than migrate in. The net migration factor is determined from the simulated data. It is different for different types of tags due to the quantum correlations which alter the distribution of the $K_S^0 K \pi$ decay on the Dalitz plot. The net migration factor is treated in the same way as a systematic uncertainty on the efficiency in the analysis. The net migration factor, that is the net loss of events inside the region, varies between tag categories and is of order 1$\%$ of the yield within the region.

Tables~\ref{tab:kskappapinfo}--\ref{tab:valkskappaminfo} give the results for the observable $\kappa$ and its uncertainties for the $K^{*}(892)^\pm$ bin. The average values are $\kappa^+$ = 0.44 $\pm$ 0.30 and $\kappa^-$ = 3.18 $\pm$ 1.08 with a $\chi^2$/dof of 6.5/10. The uncertainties due to non--uniform efficiency are 0.002 and 0.001 for $p$ and $q$, respectively. They are reduced in comparison to the full Dalitz plot measurement as the efficiency is more uniform over the restricted region.

\begin{table*}[thpb]
\begin{center}
\caption{Systematic uncertainties on the $\kappa^+$ values determined in the restricted region.\label{tab:kskappapinfo}}
\begin{tabular}{lcccccc}
\hline \hline
& $KK$  & $\pi^+\pi^-$  & $K_L^0\pi^0$  & $K_L^0\omega$  & $K_L^0\eta$  & $K_S^0\pi^0\pi^0$ \\
\hline
Tag branching fraction & 0.004 & 0.004 & 0.109 & 0.058 & 0.045 & 0.149\\
 $D^0 	\to K\pi$ branching fraction & 0.003 & 0.002 & 0.021 & 0.005 & 0.005 & 0.011\\
Background scaling ratio & 0.091 & 0.230 & 0.062 & 0.032 & 0.126 & 0.017\\
$K\pi$ tag yield& 0.042 & 0.039 & 0.338 & 0.078 & 0.082 & 0.177\\
$CP$ tag tracking and reconstruction & 0.002 & 0.001 & 0.033 & 0.008 & 0.000 & 0.019\\
$K\pi$ tracking and reconstruction& 0.001 & 0.001 & 0.011 & 0.003 & 0.003 & 0.006\\
Bin migration & 0.000 & 0.000 & 0.003 & 0.001 & 0.001 & 0.001\\
\\
Total syst. & 0.100 & 0.233 & 0.363 & 0.103 & 0.157 & 0.233\\
\hline
\hline
\end{tabular}
\end{center}
\end{table*}

\begin{table*}[thpb]
\begin{center}
\caption{The central values of $\kappa^+$ determined in the restricted region and the statistical and systematic uncertainties.\label{tab:valkskappapinfo}}
\begin{tabular}{lcccccc}
\hline \hline
& $KK$  & $\pi^+\pi^-$  & $K_L^0\pi^0$  & $K_L^0\omega$  & $K_L^0\eta$  & $K_S^0\pi^0\pi^0$ \\
\hline
Central $\kappa$ value & 0.205 & 0.192 & 1.665 & 0.386 & $-0.402$ & 0.873\\
\\
Statistical uncertainty & 0.491 & 1.059 & 0.532 & 0.579 & 0.730 & 0.755\\
Total syst. & 0.100 & 0.233 & 0.363 & 0.103 & 0.157 & 0.233\\
\\
Total stat. + syst. & 0.501 & 1.085 & 0.644 & 0.588 & 0.746 & 0.790\\
\hline
\hline
\end{tabular}
\end{center}
\end{table*}

\begin{table*}[!htbp]
\begin{center}
\caption{Systematic uncertainties on the $\kappa^-$ values determined in the restricted region.\label{tab:kskappaminfo}}
\begin{tabular}{lcccccc}
\hline \hline
 & $K_S^0\pi^0$  & $K_S^0\omega$  & $K_S^0\eta ( \gamma\gamma )$  & $K_S^0\eta(\pi\pi\pi^0)$  & $K_S^0\eta '$  & $K_L^0\pi^0\pi^0$ \\
\hline
Tag branching fraction & 0.188 & 0.656 & 0.665 & 0.266 & 1.339 & 0.432\\
 $D^0 	\to K\pi$ branching fraction & 0.051 & 0.056 & 0.077 & 0.031 & 0.133 & 0.033\\
Background scaling ratio & 0.004 & 0.008 & 0.016 & 0.045 & 0.000 & 0.054\\
$K\pi$ tag yield& 0.812 & 0.884 & 1.212 & 0.483 & 2.094 & 0.515\\
$CP$ tag tracking and reconstruction & 0.086 & 0.095 & 0.054 & 0.052 & 0.102 & 0.050\\
$K\pi$ tag tracking and reconstruction & 0.027 & 0.029 & 0.040 & 0.016 & 0.069 & 0.017\\
Bin migration & 0.006 & 0.005 & 0.007 & 0.003 & 0.011 & 0.003\\

\\
Total syst. & 0.840 & 1.107 & 1.386 & 0.557 & 2.493 & 0.677\\

\hline
\hline
\end{tabular}
\end{center}
\end{table*}
\begin{table*}[!htbp]
\begin{center}
\caption{The central values of $\kappa^-$ determined in the restricted region and the statistical and systematic uncertainties.\label{tab:valkskappaminfo}}
\begin{tabular}{lcccccc}
\hline \hline
 & $K_S^0\pi^0$  & $K_S^0\omega$  & $K_S^0\eta ( \gamma\gamma )$  & $K_S^0\eta(\pi\pi\pi^0)$  & $K_S^0\eta '$  & $K_L^0\pi^0\pi^0$ \\
\hline
Central $\kappa$ value & 4.003 & 4.357 & 5.975 & 2.382 & 10.324 & 2.537\\
\\
Statistical uncertainty & 0.994 & 1.498 & 3.523 & 4.269 & 10.881 & 2.640\\
Total syst. & 0.840 & 1.107 & 1.386 & 0.557 & 2.493 & 0.677\\
\\
Total stat. + syst. & 1.301 & 1.862 & 3.786 & 4.305 & 11.163 & 2.726\\
\hline
\hline
\end{tabular}
\end{center}
\end{table*}

\begin{figure}[!htbp]
\begin{center}
{\includegraphics[width=0.99\columnwidth]{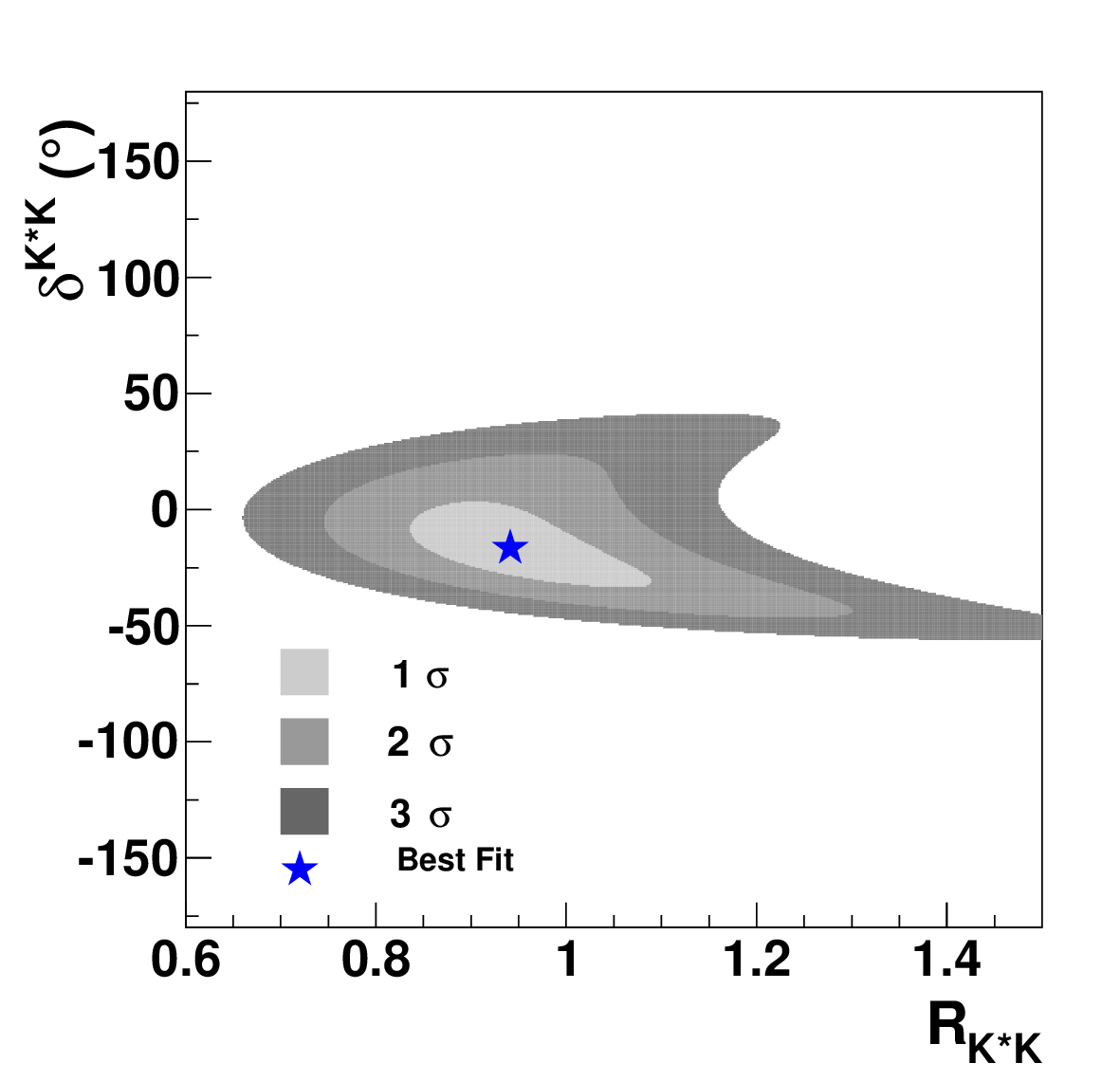}}
\caption{The best--fit point for \chokst\ and \spdkst\ which are measured over the restricted region, and the regions enclosing 1, 2, and 3 standard deviations from that point.\label{fig:cont2}}
\end{center}
\end{figure}

The fitted values of $p$ and $q$ are $p=0.902\pm 0.086$ and $q=-0.269\pm 0.314$, with $\chi^2$/dof given by 45.3/40. This leads to a measurement of \chokst$ = 0.94\pm0.12$ and \spdkst$ = (-16.6\pm18.4)^\circ$. The regions consistent with one, two, and three standard deviations from the best--fit point are shown in ~\figref{fig:cont2}. The coherence in the region of the $K^*(892)^\pm$ is larger than the coherence over the full Dalitz plot and is consistent with one. The prediction from the isobar models for the restricted region is 0.93 (0.94) when favored and suppressed models 1 (2) are used. The predictions are consistent with the measured value. 

\subsection{Impact on $\gamma$ measurements}

It is instructive to estimate the impact of the knowledge of the coherence factor and average strong--phase difference on a measurement of $\gamma$ performed at LHCb or a future high luminosity $e^+e^-$ experiment. Simple simulated data are generated assuming 15000 $B^- \to \widetilde{D} K^-$ events with $\widetilde{D}$ decaying to $K_S^0K^\pm\pi^\mp$, and  $\gamma$ = 70$^\circ$, $\delta_B$ = 130$^\circ$ and $r_B$ = 0.1, and $r_D^{K_S^0K\pi}$ and \cho\ and \spd\ as measured here for the full Dalitz plot. The chosen values for $\gamma$, $\delta_B$, and $r_B$ are consistent with current knowledge~\cite{PDG}. The data samples are fit with only the parameter $\gamma$ free. The width of the best--fit $\gamma$ distribution is 10$^\circ$. The restricted region will have approximately 60$\%$ of the statistics of the full Dalitz plot. A similar study in the restricted region assuming 9000 events and the values of $r_D^{K^*K}$ and \chokst\ and \spdkst\ as measured in the restricted region has a best--fit $\gamma$ distribution of 5$^\circ$. The better sensitivity, despite the smaller sample, is due to a combination of increased coherence, a lower value of  $r_D^{K*K}$ in comparison to $r_D^{K_S^0K\pi}$, and a different value of the average strong--phase difference. The uncertainties on the coherence factor and average strong--phase difference are incorporated by making them free parameters in the fit and applying Gaussian constraints based on their measured uncertainties. In this case the width of the best--fit $\gamma$ distribution increases to 14$^\circ$ and 7$^\circ$ for the full Dalitz plot and restricted regions, respectively. The uncertainties \cho, \spd, \chokst, and \spdkst\ are dominated by the statistical uncertainties and hence the precision of any measurement of $\gamma$ could be increased with a more accurate measurement of the coherence factors and average strong--phase differences. The inclusion of this decay channel in a global fit to $\gamma$ with other channels will improve the overall sensitivity to this parameter.

\section{Summary}
\label{sec:sum}

Isobar models for the decays \dtofav\ and \dtosup\ have been presented. They are dominated by the $K^*(892)^\pm$ resonance in the case of the favored decay and by the same resonance and a broad S--wave component in the suppressed decay.  The ratio of the branching fractions has been measured, and is more precise that the previous average~\cite{PDG}. The coherence factor and average strong--phase difference for $D^0 \to K_S^0 K\pi$ have been measured in two regions. The first covers all final state kinematics, and the second covers a restricted region around the $K^{*}(892)^\pm$ resonance. The coherence is high, which makes the $D^0 \to K_S^0 K\pi$ channel interesting for future determinations of the CKM phase $\gamma$. The measured values of \braf, \cho, and \spd\ are also all useful inputs for charm mixing measurements as described in ~\cite{dmix}.

\begin{acknowledgments}
We gratefully acknowledge the effort of the CESR staff
in providing us with excellent luminosity and running conditions.
This work was supported by
the National Science Foundation,
the U.S. Department of Energy,
the Natural Sciences and Engineering Research Council of Canada, and
the U.K. Science and Technology Facilities Council.
\end{acknowledgments}

\appendix
\section{The interference fit fractions}
This appendix gives the full interference fit fractions as defined in ~\equationref{eq:interference} between each resonance for each model in Tables ~\ref{tab:interference1}--\ref{tab:interference4}.

\begin{table*}[!htbp]
\caption{ The interference fit fractions between every pair of resonances in favored model 1. The fractions are provided as a percent.\label{tab:interference1}}
\begin{tabular}[c]{llllll}
\hline\hline
  & $K^*(892)^0$ & $K^*(1410)^0$ & $K^*(1680)^-$ & $K_0^*(1430)^-$ & $a_0(1450)^+$\\
\hline
$K^*(892)^-$ & $-$0.23$\pm$0.65$\pm$0.57 & $-$12.29$\pm$1.23$\pm$3.42 & 13.62$\pm$3.32$\pm$2.17 & 3.38$\pm$0.88$\pm$0.63 & 2.56$\pm$2.09$\pm$1.71\\
$K^*(892)^0$ &  & $-$2.81$\pm$1.55$\pm$0.79 & 0.23$\pm$0.27$\pm$0.75 & 0.57$\pm$0.30$\pm$0.24 & 0.26$\pm$0.30$\pm$0.12\\
$K^*(1410)^0$ &  &  & $-$21.18$\pm$2.73$\pm$8.66 & 1.15$\pm$0.61$\pm$1.39 & 0.78$\pm$3.19$\pm$1.99\\
$K^*(1680)^-$ &  &  &  & $-$6.70$\pm$1.79$\pm$2.68 & 0.55$\pm$4.08$\pm$2.00\\
$K_0^*(1430)^-$ &  &  &  &  & 1.37$\pm$1.17$\pm$0.75\\
\hline\hline
\end{tabular}
\end{table*}

\begin{table*}[!htbp]
\caption{ The interference fit fractions between every pair of resonances in favored model 2. The fractions are provided as a percent.\label{tab:interference2}}
\begin{tabular}[c]{llllll}
\hline\hline
  & $K^*(1410)^0$ & $K^*(1680)^-$ & $\rho(1700)^+$ & $K_0^*(1430)^-$ & $a_0(1450)^+$\\
\hline
$K^*(892)^-$ & $-$14.04$\pm$1.42$\pm$2.02 & 13.05$\pm$2.61$\pm$4.02 & 3.60$\pm$1.52$\pm$3.36 & 3.58$\pm$0.70$\pm$0.92 & 4.03$\pm$1.60$\pm$1.52\\
$K^*(1410)^0$ &  & $-$27.47$\pm$5.29$\pm$28.07 & $-$7.33$\pm$3.45$\pm$8.73 & 1.55$\pm$0.74$\pm$2.30 & 1.82$\pm$3.41$\pm$1.72\\
$K^*(1680)^-$ &  &  & $-$4.33$\pm$2.76$\pm$7.87 & $-$8.06$\pm$1.93$\pm$7.16 & $-$0.30$\pm$4.16$\pm$1.98\\
$\rho(1700)^+$ &  &  &  & 1.24$\pm$0.55$\pm$1.57 & 0.02$\pm$0.01$\pm$0.01\\
$K_0^*(1430)^-$ &  &  &  &  & 2.49$\pm$1.34$\pm$0.86\\
\hline\hline
\end{tabular}
\end{table*}

\begin{table*}[ht]
\caption{ The interference fit fractions between every pair of resonances in suppressed model 1. The fractions are provided as a percent.\label{tab:interference3}}
\begin{tabular}[c]{lllll}
\hline\hline
  & $K^*(892)^0$ & $K_0^*(1430)^-$ & $a_0(1450)^+$ & $\rho(1700)^+$\\
\hline
$K^*(892)^-$ & $-$0.74$\pm$0.46$\pm$0.10 & 4.91$\pm$0.57$\pm$0.53 & 7.01$\pm$0.88$\pm$0.87 & 5.74$\pm$0.91$\pm$1.16\\
$K^*(892)^0$ &  & 1.34$\pm$0.54$\pm$0.33 & $-$0.50$\pm$0.58$\pm$0.40 & 0.72$\pm$0.47$\pm$0.29\\
$K_0^*(1430)^-$ &  &  & 17.39$\pm$1.52$\pm$1.69 & $-$0.49$\pm$0.75$\pm$0.41\\
$a_0(1450)^+$ &  &  &  & $-$0.01$\pm$0.01$\pm$0.01\\
\hline\hline
\end{tabular}
\end{table*}

\begin{table*}[!htbp]
\caption{ The interference fit fractions between every pair of resonances in suppressed model 2. The fractions are provided as a percent.\ \label{tab:interference4}}
\begin{tabular}[c]{lllll}
\hline\hline
  & $K_0^*(1430)^-$ & $\rho(1700)^+$ & $a_0(1450)^+$ & $K_2^*(1430)^0$\\
\hline
$K^*(892)^-$ & 5.70$\pm$0.55$\pm$0.31 & 6.41$\pm$0.89$\pm$1.41 & 3.62$\pm$0.95$\pm$0.64 & 1.75$\pm$0.81$\pm$3.32\\
$K_0^*(1430)^-$ &  & 2.07$\pm$0.85$\pm$0.80 & 15.25$\pm$2.21$\pm$0.91 & $-$5.79$\pm$5.60$\pm$9.15\\
$\rho(1700)^+$ &  &  & 0.00$\pm$0.01$\pm$0.01 & 4.00$\pm$1.23$\pm$1.71\\
$a_0(1450)^+$ &  &  &  & $-$3.71$\pm$1.64$\pm$2.26\\
\hline\hline
\end{tabular}
\end{table*}

\section{The covariance matrices}

Tables ~\ref{tab:covar} and ~\ref{tab:covarKbin} give the the covariance matrices for the uncertainties between the $\kappa$ measurements from the different tags. Entries below the diagonal are not shown.
\begin{sidewaystable}
\begin{center}
\caption{The covariance matrix for the $\kappa$ observables determined using data from the full Dalitz plot. \label{tab:covar}}
\begin{small}
\begin{tabular}{lcccccccccccccccc}
\hline \hline

 & $\kappa^{KK}$  & $\kappa^{\pi\pi}$  & $\kappa^{K_L\pi^0}$  & $\kappa^{K_L\omega}$  & $\kappa^{K_L\eta( \gamma\gamma) }$  & $\kappa^{K_L\eta '}$  & $\kappa^{K_S\pi^0\pi^0}$  & $\kappa^{K_S\pi^0}$  & $\kappa^{K_S\omega}$  & $\kappa^{K_S\eta ( \gamma\gamma )}$  & $\kappa^{K_S\eta ( \pi \pi \pi^0 )}$  & $\kappa^{K_S\eta '}$  & $\kappa^{K_L\pi^0\pi^0}$ \\
$\kappa^{KK}$ & 0.11 & 0.01 & 0.01 & 0.00 & 0.01 & 0.01 & 0.01 & 0.01 & 0.01 & 0.02 & 0.01 & 0.01 & 0.01\\
$\kappa^{\pi\pi}$ &  & 0.78 & 0.03 & 0.01 & 0.03 & 0.02 & 0.05 & 0.05 & 0.06 & 0.10 & 0.02 & 0.06 & 0.06\\
$\kappa^{K_L\pi^0}$ &  &  & 0.19 & 0.01 & 0.01 & 0.02 & 0.08 & 0.07 & 0.09 & 0.16 & 0.03 & 0.09 & 0.07\\
$\kappa^{K_L\omega}$ &  &  &  & 0.23 & 0.00 & 0.01 & 0.03 & 0.03 & 0.08 & 0.06 & 0.01 & 0.03 & 0.03\\
$\kappa^{K_L\eta( \gamma\gamma) }$ &  &  &  &  & 0.46 & 0.01 & 0.00 & 0.00 & 0.00 & 0.01 & 0.01 & 0.00 & 0.01\\
$\kappa^{K_L\eta '}$ &  &  &  &  &  & 2.91 & 0.03 & 0.02 & 0.03 & 0.05 & 0.01 & 0.06 & 0.03\\
$\kappa^{K_S\pi^0\pi^0}$ &  &  &  &  &  &  & 1.61 & 0.16 & 0.20 & 0.36 & 0.05 & 0.20 & 0.43\\
$\kappa^{K_S\pi^0}$ &  &  &  &  &  &  &  & 0.63 & 0.19 & 0.35 & 0.05 & 0.19 & 0.15\\
$\kappa^{K_S\omega}$ &  &  &  &  &  &  &  &  & 1.91 & 0.42 & 0.06 & 0.23 & 0.18\\
$\kappa^{K_S\eta ( \gamma\gamma )}$ &  &  &  &  &  &  &  &  &  & 8.53 & 0.21 & 0.43 & 0.33\\
$\kappa^{K_S\eta ( \pi \pi \pi^0 )}$ &  &  &  &  &  &  &  &  &  &  & 6.77 & 0.06 & 0.05\\
$\kappa^{K_S\eta '}$ &  &  &  &  &  &  &  &  &  &  &  & 21.78 & 0.18\\
$\kappa^{K_L\pi^0\pi^0}$ &  &  &  &  &  &  &  &  &  &  &  &  & 4.86\\

\hline
\hline
\end{tabular}
\end{small}
\end{center}
\end{sidewaystable}

\begin{sidewaystable}
\begin{center}
\caption{The covariance matrix for the $\kappa$ observables determined using data from the restricted region. \label{tab:covarKbin}}
\begin{small}
\begin{tabular}{lcccccccccccccccc}
\hline \hline

 & $\kappa^{KK}$  & $\kappa^{\pi\pi}$  & $\kappa^{K_L\pi^0}$  & $\kappa^{K_L\omega}$  & $\kappa^{K_L\eta( \gamma\gamma) }$  & $\kappa^{K_S\pi^0\pi^0}$  & $\kappa^{K_S\pi^0}$  & $\kappa^{K_S\omega}$  & $\kappa^{K_S\eta ( \gamma\gamma )}$  & $\kappa^{K_S\eta ( \pi \pi \pi^0 )}$  & $\kappa^{K_S\eta '}$  & $\kappa^{K_L\pi^0\pi^0}$ \\
$\kappa^{KK}$ & 0.25 & 0.02 & 0.02 & 0.01 & 0.01 & 0.01 & 0.03 & 0.04 & 0.05 & 0.02 & 0.09 & 0.03\\
$\kappa^{\pi\pi}$ &  & 1.18 & 0.03 & 0.01 & 0.03 & 0.01 & 0.03 & 0.04 & 0.05 & 0.03 & 0.08 & 0.03\\
$\kappa^{K_L\pi^0}$ &  &  & 0.41 & 0.03 & 0.04 & 0.06 & 0.28 & 0.30 & 0.41 & 0.17 & 0.71 & 0.18\\
$\kappa^{K_L\omega}$ &  &  &  & 0.35 & 0.01 & 0.01 & 0.06 & 0.11 & 0.10 & 0.04 & 0.16 & 0.04\\
$\kappa^{K_L\eta( \gamma\gamma) }$ &  &  &  &  & 0.56 & 0.02 & 0.07 & 0.07 & 0.13 & 0.06 & 0.17 & 0.05\\
$\kappa^{K_S\pi^0\pi^0}$ &  &  &  &  &  & 0.62 & 0.14 & 0.16 & 0.22 & 0.09 & 0.37 & 0.16\\
$\kappa^{K_S\pi^0}$ &  &  &  &  &  &  & 1.69 & 0.72 & 0.99 & 0.39 & 1.71 & 0.42\\
$\kappa^{K_S\omega}$ &  &  &  &  &  &  &  & 3.47 & 1.08 & 0.43 & 1.86 & 0.46\\
$\kappa^{K_S\eta ( \gamma\gamma )}$ &  &  &  &  &  &  &  &  & 14.33 & 0.77 & 2.55 & 0.63\\
$\kappa^{K_S\eta ( \pi \pi \pi^0 )}$ &  &  &  &  &  &  &  &  &  & 18.53 & 1.02 & 0.25\\
$\kappa^{K_S\eta '}$ &  &  &  &  &  &  &  &  &  &  & 124.62 & 1.08\\
$\kappa^{K_L\pi^0\pi^0}$ &  &  &  &  &  &  &  &  &  &  &  & 7.43\\
\hline
\hline
\end{tabular}
\end{small}
\end{center}
\end{sidewaystable}

\nopagebreak

\end{document}